\documentclass[11pt]{report}
\usepackage{epsfig}
\begin{document}

\pagestyle{plain}

\pagenumbering{roman}

\tableofcontents


\chapter*{Abstract}
\addcontentsline{toc}{chapter}{Abstract}
\noindent
This work is related to different questions within cosmology. 
The principal idea herein is to develop cosmological
knowledge making use of the analyses of observational data in order to 
find the values of the matter density $\Omega_m$ and vacuum energy density 
$\Omega_\Lambda$. 

Data fitting is carried out using two statistical 
methods, $\chi^2$ and maximum likelihood. The data analysis 
exhibits that a low density and flat Universe is strongly favoured.  

Applying the $\Omega_m$ value found for clusters of galaxies, we 
demonstrate that clusters have very little room for baryonic dark matter.  
An upper limit to the small but non-negligible sum of 
baryonic dark matter and galaxy mass can be estimated, requiring the 
use of special statistics.  

A Toroidal Black Hole (TBH) study, in contrast to the Spherical Black
Hole (SBH), shows that the TBH can be used as an important 
tool in explaining AGN phenomena. 

\newpage

\pagenumbering{arabic}

\chapter{Cosmological Models and Parameters}

Measurement of global cosmological parameters is one of the key challenges 
in current cosmology research. People are interested in knowing in what 
kind of Universe we live. Only a precise measurement of density parameters 
can tell us the critical nature of the Universe. The parameters determined 
from the different observations are strongly model dependent and also 
contradictory to each other. Resolution of contradiction is another big 
challenge to the scientist. A lot of interesting discussions is reported  
in this thesis, based on recent output from the cosmological research. 

According to the cosmological principle, the Universe is 
homogeneous and isotropic. Most of the cosmological models are based on 
Friedman equations, describing a Universe in expansion or contraction. 
Einstein sought a solution in which the Universe would be static and 
eternal because the Universe known at that time comprised only the Milky 
Way, which clearly was not expanding or contracting. In this Chapter a few 
models will be discussed that are remarkable in modern cosmology.  

\begin{figure}[t]
\centering
\mbox{\epsfxsize=10.7truecm\epsfysize=9.3truecm\epsffile{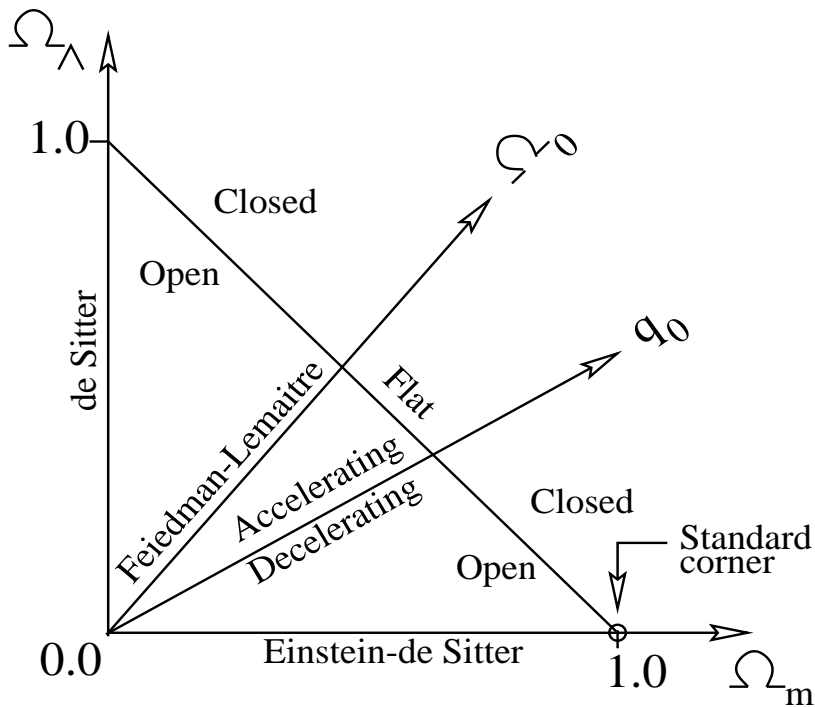}}
\caption{Geometry of different types of Universe in $(\Omega_m,\ 
\Omega_\Lambda)$ plane. Flat and open models are infinite; closed models
are finite. There are also possibilities that an infinite Universe can 
collapse if $\Omega_\Lambda < 0$. A finite Universe expands for ever
if $\Omega_\Lambda = 0$ with $\Omega_m \leq 1$, and collapses into a big
crunch if $\Omega_m > 1$. A Finite Universe can be accelerating/decelerating.} 
  \label{fig1}
\end{figure}

This thesis deals with {\it Cosmological Parameters} and {\it Black Holes.} 
Gravitational matter $\Omega_m$ is composed of baryonic matter, baryonic  
dark matter and dark matter of unknown composition. The matter density
($\Omega_m$), baryonic density ($\Omega_b$), Hubble constant ($H_0$) and  
the age of the Universe ($t_0$) will also be discussed. A discussion on the  
cosmological constant ($\Lambda$) and its vacuum energy density 
($\Omega_\Lambda$) will be presented in the next Chapter. 
A simple picture of different cosmologies are portrayed in Fig. \ref{fig1}. 
Some aspects of {\it Black Hole} physics will be discussed in Chapter 6. 

\section{Einstein's Model}  

Modern 
cosmological models began with Einstein's static cosmology. Einstein proposed
that the gravity is not a force but merely a manifestation of free motion in
space-time. In addition, the geometry of space-time is determined by the 
energy in the Universe. The energy is associated with mass, radiation and 
pressure. A cosmological constant $\Lambda$ was introduced 
by Einstein into the equation of General Relativity to allow for a stationary  
solution, and it was introduced before the concept of Big Bang. 
However, the idea of a static Universe fails to agree with 
observations, because the Universe is expanding linearly with time.  

\section{The de Sitter Model} 

In 1917, Willem de Sitter discovered a cosmological model differing 
from Einstein's. He observed that the light from distant objects becomes 
redder as the distance increases. There is no matter in this model. The 
space-time is flat, and the space expands exponentially because of the  
$\Lambda$ force. This is one of the earliest models that was considered 
when the modern science of cosmology was in its infancy. 

\section{The Einstein-de Sitter Model}

According to Einstein and de Sitter, the cosmological constant should be set
equal to zero, and they derived a homogeneous and isotropic model.
They assumed that the spatial curvature of the Universe
is neither positive nor negative but rather zero. The special geometry 
of the Einstein-de Sitter Universe is Euclidean, but the space-time is not
globally flat. People with a philosophical bent have long considered 
it as the most fitting candidate to describe the actual Universe. Strong 
theoretical support for this viewpoint came from particle physics, but as 
yet not definitive, astronomical observations also supported this model.

\section{The Friedman-Lema\^{\i}tre Model}

The standard model of cosmology describes the Universe as expanding 
at present.
The formulation and prediction of a Big Bang expansion for the Universe
is remarkable. Both Friedman and Lema\^{\i}tre in different 
years, independently discovered the
solution to Einstein's equations of gravitation which described an 
expanding Universe.  The expansion could either continue forever, or
eventually reverse into a phase of contraction.

However, the Friedman-Lema\^{\i}tre evolutionary model could not 
predict explicitly the curvature of the expansion, permitting three
possible solutions: either hyperbolic, flat or spherical expansion.
Friedman's equations may be written (Kolb \& Turner 1990, Roos 1997, 
Bergstr$\rm\ddot{o}$m \& Goober 1999) as
\begin{eqnarray}
{\dot{R}^2 + kc^2 \over R^2} - {\Lambda \over 3} = {8 \pi G \rho \over 3} 
\label{f11}\end{eqnarray}
\begin{eqnarray}
{2\ddot{R} \over R} + {\dot{R}^2 + kc^2 \over R^2}  - \Lambda = - {8 \pi G p \over c^2} , 
\label{f12}\end{eqnarray}
where {\it R} is the scale factor of the Universe, {\it k} is the curvature 
parameter, $\Lambda$ is the cosmological constant, {\it G} is the Newtonian 
constant, $\rho$ is the total energy density and {\it p} is the total pressure.

There is a relation between pressure and density, called equation of state
\begin{eqnarray}
p=\alpha \cdot \rho
\label{f13}\end{eqnarray}
where $\alpha$ is a constant, different in different eras of
the Universe. In the radiation dominated era {\it $\rho\sim {1 \over R^4}$}, 
and in the matter dominated era {\it $\rho\sim {1 \over R^3}$}. For the 
de Sitter model $\alpha$ = -1, but for interacting fields 
and topological defects $\alpha$ can vary between -1 and 0. Since recent 
supernovae observations show that the Universe is accelerating, one has
-1 $\leq$ $\alpha$ $<$ -${1\over 3}$. Models with $\alpha$ in this range  
some people call {\it Quintessence models}.

The critical density of the Universe at redshift {\it z} is 

\begin{eqnarray}
\rho_c(z) = {3H^2(z) \over 8 \pi G}
\label{f14}\end{eqnarray}
where {\it H} is the Hubble constant.
The vacuum energy density parameter $\Omega_\Lambda$ is related to the 
cosmological constant $\Lambda$ by 
\begin{eqnarray}
\Omega_\Lambda = {\Lambda \over 3H_0^2} = {\rho_\Lambda \over \rho_c} ,
\label{f15}\end{eqnarray}
where the notation zero means the value at present.  
The matter density parameter of the Universe is 
\begin{eqnarray}
\Omega_m = {\rho_m \over \rho_c} .
\label{f16}\end{eqnarray}

The total density parameter of the Universe is the sum of matter density 
parameter and the vacuum energy density parameter
\begin{eqnarray}
\Omega_0 = \Omega_m + \Omega_\Lambda . 
\label{f17}\end{eqnarray}

The flat condition of the Universe is  [{\it k} = 0 in Eqs. (\ref{f11}) and 
(\ref{f12})]
\begin{eqnarray}
\Omega_m + \Omega_\Lambda = 1
\label{f18}\end{eqnarray}

We can arrive at a relation among $H_0, \Omega_m$, $\Omega_{\Lambda}$ 
and $t_0$, from the Friedman-Lema\^{\i}tre model, as
\begin{eqnarray}
t_0={{1}\over{H_0}}\int_0^1\hbox{d}x
\big{[}(1-\Omega_m-\Omega_{\Lambda})+\Omega_mx^{-1}+\Omega_{\Lambda}x^2
\big{]}^{-1/2}.  
\label{f19}\end{eqnarray}

This equation will be used to determine the age of  
the Universe. There is no analytical solution to this integration but 
it can easily be evaluated
numerically. To determine $t_0$, we need to have precise values of other
parameters. The evaluation of $\Omega_m$, $\Omega_{\Lambda}$ can be found
in Papers (II), (III), (IV) and also in the next Chapters.

\section{Density Parameters}

Generally, matter density and vacuum energy density are treated as Universal 
density parameters, and the total matter density is the combination of 
different kinds of matter. So far as we know, there are two kinds of matter 
in the Universe, baryonic matter and dark matter. Thus the total  
mass density parameter $\Omega_m$ is the sum of these two. 
The candidates of dark matter and their nature will be discussed 
in the Chapter 5. For a long time, the physicists had  
neglected $\Omega_\Lambda$, simply it was considered that 
$\Omega_\Lambda = 0$ and $\Omega_m = 1$. The situation has changed in the 
last few years by the enormous data collection of powerful telescopes and 
by the success of satellite missions. Observations suggest that this is 
not our real Universe.

Due to the expansion of the Universe the mass density is decreasing and 
the vacuum energy density is now dominating. The vacuum energy is related 
to the cosmological constant, Eq. (\ref{f15}).  
Most recent measurements agree 
that the value of present matter density $\Omega_m \sim {1\over3}$ in the 
flat case. It is the proper time for the astrophysicists and cosmologists to   
have a direct constraint on the density parameters. 
The constraints on matter density $\Omega_m$ and vacuum energy 
density $\Omega_\Lambda$ are extensively described in the next Chapters.

\section{Baryonic Density}

The value of the universal baryonic density parameter $\Omega_b$ follows 
from standard Big Bang Nucleosynthesis (BBN) arguments, 
and from the observed abundances of $^4$He, D, $^3$He, and $^7$Li (cf. eg. 
Sarkar 1999), in particular from the low deuterium abundance measured by 
Burles et al. (1999, 2001). Their estimate is
\begin{eqnarray}
\Omega_b h^2 = 0.020\pm 0.001\  (68\% \ {\rm CL}) 
\label{Omega_b1}\end{eqnarray} 
where $H_0 = 100h$. Thus, taking this central value and the Hubble constant 
from Gibson \& Brook (2001) we get today's 
universal baryonic density $\sim 4\%$. The BBN value is in disagreement with 
the recent observations of CMB anisotropy which yields (Jaffe et al. 2001)   
\begin{eqnarray}
\Omega_b h^2 = 0.033_{-0.004}^{+0.005}\ (68\% \ {\rm CL}). 
\label{Omega_b2}\end{eqnarray}
Note that BBN provides a probe of the 
Universal abundance of baryons when the Universe was only a few minutes old. 
Observations of CMB anisotropy probe the baryon abundance when the Universe
was 3-5 hundred thousand years old, and SNe Ia supernovae and clusters of 
galaxies observations probe a more recent past, when the Universe was 
several billion years old. 

Most recently, the Degree Angular 
Scale Interferometer (DASI) has measured the angular power spectrum of 
the Cosmic Microwave Background anisotropy over the range 100 $<$ {\it l} 
$>$ 900 (Pryke et al. 2001). Here, the second peak is more pronounced than 
found BOOMERANG (de Bernardis et al. 2000) and MAXIMA-1 (Balbi et al. 2000), 
and the contradiction with BBN can be resolved. DASI (Pryke et al. 2001) 
gives a new value of baryonic density as     
\begin{eqnarray} 
\Omega_b h^2 = 0.022_{-0.003}^{+0.004}\ (68\% \ {\rm CL}), 
\label{Omega_b3}\end{eqnarray} 
which is a good agreement with BBN (Burles et al. 2001). DASI has also 
independently determined $\Omega_m = 0.40 \pm 0.15$ and 
$\Omega_\Lambda = 0.60 \pm 0.15$ (68\% CL), although the errors ber is large  
and no contour plot is given. Adding the data at higher $l$ BOOMERANG 
(de Bernardis 2001) has reported the value of $\Omega_b h^2$, their new 
value is exactly same as (DASI 2001) Eq. (\ref{Omega_b3}).   
 
Assuming that the BBN value for $\Omega_b h^2$ in Eq. (\ref{Omega_b1}) 
(Burles et al. 2001) is 
correct, and taking the value of the Hubble constant $ h = 0.73 \pm 0.05$ 
Gibson \& Brook (2001), one can obtains a constraint on the baryonic density 
parameters, $\Omega_b = 0.038 \pm 0.002$.

\section{The Hubble Constant} 

In 1929, Edwin Hubble discovered that the distant galaxies are 
moving away from us. A simple mathematical relation between the 
Hubble constant and distant objects 
is $H_0 = {v/d}$, where {\it v} is the galaxy's radial velocity and {\it d} 
is the galaxy's distance from the earth.
This is one of the important parameter because it is directly related to the 
age of the Universe, and it is also related to the other cosmological 
parameters. The Hubble constant is well determined from the 
HST Key Project on the Extragalactic Distance Scale (Mould et al. 2000), 
obtained by the combination of four methods (SBF, FP, Tully-Fisher, and 
SNe Ia). Their combined result is $H_0 = 71 \pm 6 
{\rm \ km \ s^{-1} Mpc^{-1}}$. 
A new value of the Hubble constant is presented by Gibson \& Brook (2001) 
performing the re-analysis of the combined Calan-Tololo 
and Center for Astrophysics 
(CfA) type SNe Ia datasets. The fit is extremely good with 3\% statistical 
and 10\% systematic errors. With a calibration to the corrected 
Hubble Diagrams by seven high-quality nearby SNe Ia the Hubble constant
becomes
\begin{eqnarray}
H_0 = 73 \pm 2(r) \pm 7(s) \rm{\ km \ s^{-1} Mpc^{-1}}\ .
\label{H01}\end{eqnarray}

However, the results agree with each other at $1\sigma$ level.
The Hubble constant
as measured by different methods summarized in table \ref{table4}

\begin{table}[h]
\begin{center} 
\begin{tabular}{lccc}  
\hline
{\em Method}  &{\em $H_0$} &{\em Error (\%)} & {\em References} \\ \hline
Type Ia supernovae   & 73  & $\pm{2(r)}\pm{7(s)}$  & \cite{gibs}    \\
Combined HST methods & 71 & $\pm{2(r)}\pm{6(s)}$  & \cite{free}    \\
Tully-Fisher clusters & 71 & $\pm{3(r)}\pm{7(s)}$  & \cite{saka}  \\
FP clusters & 82 & $\pm{6(r)}\pm{9(s)}$  & \cite{kels}  \\
SBF clusters & 70 & $\pm{5(r)}\pm{6(s)}$  & \cite{ferr}  \\
Type II supernovae & 72 & $\pm{9(r)}\pm{7(s)}$  & \cite{schm}  \\
Cepheids, Metallicity-corrected & 68*) & $\pm{5(r\&s)}$  & \cite{nev1}    \\
S-Z effect & 66 & $\pm{13(r)}\pm{15(s)}$  & \cite{maso}  \\
\hline  
\end{tabular}
\end{center}
\caption {The Hubble constant in units of $\rm{km \ s^{-1}Mpc^{-1}}$. 
Random (r) 
and systematic (s) errors are given at 1$\sigma$ confidence level. *) Taking 
into account more recent calibrations of the LMC distance, this value 
increases to 74 (cf. eg. Gibson \& Brook 2001). }
 \label{table4}
\end{table}

\section{Age of the Universe}

The absolute age of the Universe $t_0$ depends on 
$(h,\Omega_m,\Omega_\Lambda)$. The mass
density of the Universe $\Omega_m$ and the cosmological constant 
$\Omega_\Lambda$ are known from the present analysis. Now we can 
apply a constraint on $t_0$, since our analysis strongly preferred a flat
geometry in the Universe and this constraint is valid is the flat case 
also. In Papers II \& III we have determined the age of the Universe to be  
$t_0 = 13.5 \pm 1.3$ (0.68/h) Gyr, where we use the earlier Hubble constant 
$H_0 = 68 \pm 5 \ {\rm km \ s^{-1} Mpc^{-1}}$  
(Nevalainen \& Roos 1998). Since the age of the Universe $t_0$ strongly 
depends on the Hubble constant, our value also changes with Hubble value. 
Note that the Universe must be older than any other 
object in the Universe. Our determination agrees well 
 with others as can be seen in table \ref{table3}. 
The age determination from the different 
experiments are summarized in table \ref{table3}.  

\begin{table}[h]
\begin{center} 
\begin{tabular}{lcccc}  
\hline
{\em Technique}  & h Assumption & Age (Gyr) & Objects &{\em References} \\ 
\hline
Stellar ages & None & $13.3\pm{1.1}$ & Halo GC & \cite{jime} \\
Stellar ages & None & $12.5\pm{1.5}$ & Halo GC & \cite{grun} \\
High-Z  &  $0.65\pm{0.02} $   & $14.2\pm{1.7}$   & Universe & \cite{ries}  \\
High-Z (flat) &  $0.65\pm{0.02} $   & $15.2\pm{1.7}$ & Universe  & \cite{ries}  \\
SCP  &  0.63   & $14.2\pm{1.0}$   & Universe & \cite{per1}
    \\
SCP (flat) &  0.63  & $14.9_{-1.1}^{+1.4}$   & Universe & 
\cite{per1}    \\
4-constraints & $0.68\pm{0.10} $   & $13.4\pm{1.6}$   & Universe & 
\cite{lin2} \\
9-constraints & $0.68\pm{0.05} $   & $13.5\pm{1.3}$   & Universe & 
\cite{roo1}  \\
MCA*) & $0.70\pm{0.15} $   & $13.2_{-2.0}^{+3.6}$   & Universe & 
\cite{fer1}  \\

\hline  
\end{tabular}
\end{center}
\caption {Comparison of the age of Universe estimated using different 
observations. Errors are given at 1$\sigma$ confidence level. *) Multiband 
Colour Alalysis of bright clusters.}
 \label{table3}
\end{table}


\chapter{Problem of Missing Energy and Cosmological Constant}

\section{The Cosmological Constant}

\noindent
The cosmological constant is not a constant it might be change with time. 
It is related to vacuum energy density, Eq. (\ref{f15}), and it is a 
potentially important contributor to
the dynamical history of the Universe. The relation between the cosmological 
constant $\Lambda$ and the vacuum energy density $\Omega_\Lambda$ 
were shown in the previous Chapter. A strong evidence for a  
non-zero cosmological constant comes from the Supernovae Cosmology Project   
(SCP), Perlmutter et al. (1999), and the High Redshift Supernovae Search Team 
(HSST), Riess et al. (1998). In contrast to standard general 
relativity, a wide 
theoretical discussion on a non-zero cosmological constant can be found 
in Carroll \& Press (1992), Carroll (2000). We show that three independent 
constraints strongly rule out the standard model of flat space with vanishing 
cosmological constant (Fig. \ref{fig10} \& Fig.\ref{fig11} and for 
details Paper I). However, the present value of the cosmological constant 
is an empirical issue, a precise determination of 
which would be one of the greatest successes of observational cosmology 
in the near future. If the cosmological constant today comprises most of 
the energy density of the Universe, the age of the Universe is much 
older. 

The most surprising recent advance in cosmology is that 70\% of the Universe
seems to be made of vacuum energy. We have combined most recent observational
data in the next Chapters to demonstrate this evidence. Our two dimensional 
maximum likelihood gives a matter density of $\Omega_m = \Omega_{dark} + 
\Omega_b = 0.31$ and a vacuum energy density of $\Omega_\Lambda=0.68$. 
The total density
of the Universe is $\Omega_0=\Omega_m+\Omega_\Lambda =0.99$ 
(Paper IV). 

In figure \ref{fig10} \& Fig.\ref{fig11} we show our recent 
(Paper IV) allowed region in 
($\Omega_m, \Omega_\Lambda$)-plane. The small solid square
and the big dashed square are for the flat and two-dimensional cases 
indicated, respectively. 

\begin{figure}
\centering
\mbox{\epsfxsize=10.7truecm\epsfysize=9.3truecm\epsffile{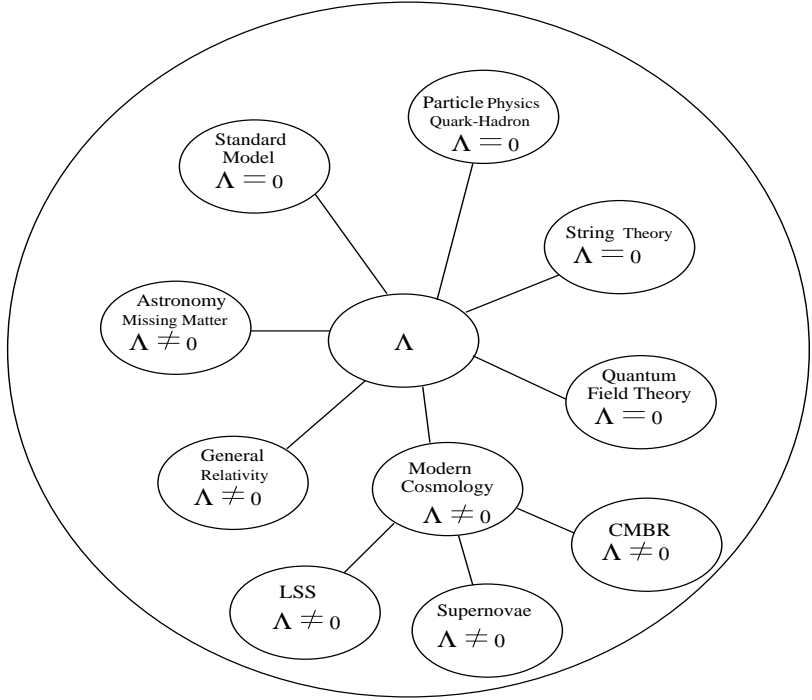}}
\caption{$\Lambda$ is in the center of physics.}
\label{fig9}
\end{figure}

\begin{figure}
\centering
\epsfig{file=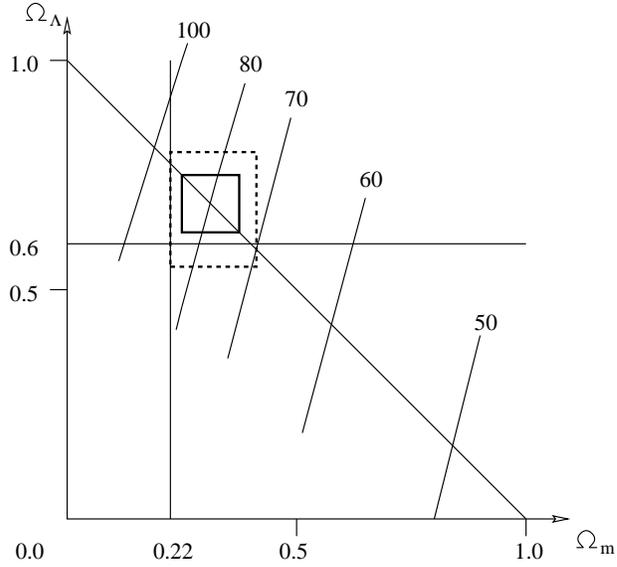,angle=0,width=0.6\textwidth}
\caption{Regions allowed in the $(\Omega_0, \Omega_{\Lambda})$-space. 
The set of solid lines are imposed by the observation of the radio galaxy 
53W091 assuming $\alpha=0$. The numbers indicate the value of the 
Hubble constant in units of km s$^{-1}$ Mpc$^{-1}$. The limits 
$\Omega_m <0.22$, $\Omega_{\Lambda}>0.6$ are indicated. The solid and dashed 
square boxes indicate the recent (Paper IV) allowed 
region in the parameter space.}
\label{fig10}
\end{figure}

\begin{figure}
\centering
\epsfig{file=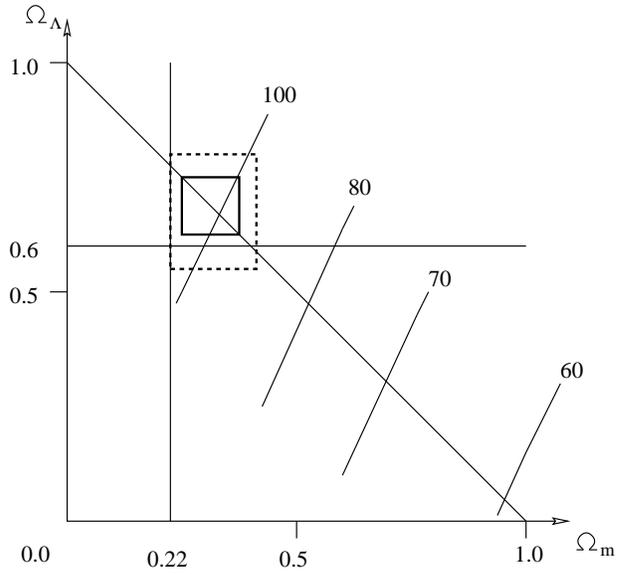,angle=0,width=0.6\textwidth}
\caption{As Fig. \ref{fig10}, but for $\alpha$ = -0.1}
\label{fig11}
\end{figure}

\section{The Missing Energy Problem} 

\noindent
The missing energy may be realized by comparing with the critical 
energy and the observed energy. Inflationary cosmology and current 
cosmic microwave anisotropy measurements suggest that the Universe 
is flat. At the same time observations indicate that the sum of 
ordinary (baryonic) and dark matter in the Universe is below the
critical density. So, the ``missing energy'' problem arises. 
It is possible that the missing energy could be 
from the interacting fields or topological defects (Steinhartz 1996, 
Huey et al. 1999, Bludman \& Roos 2001).
As the Universe expands, the missing energy density varies as 
$(R)^{-3(1+\alpha)}$, where $\alpha = 0$ for ordinary matter, $- {1\over 2} 
\ge \alpha$ $>$ - 1 for quintessence and $\alpha = - 1$ for the cosmological 
constant. 

\begin{figure}[h]
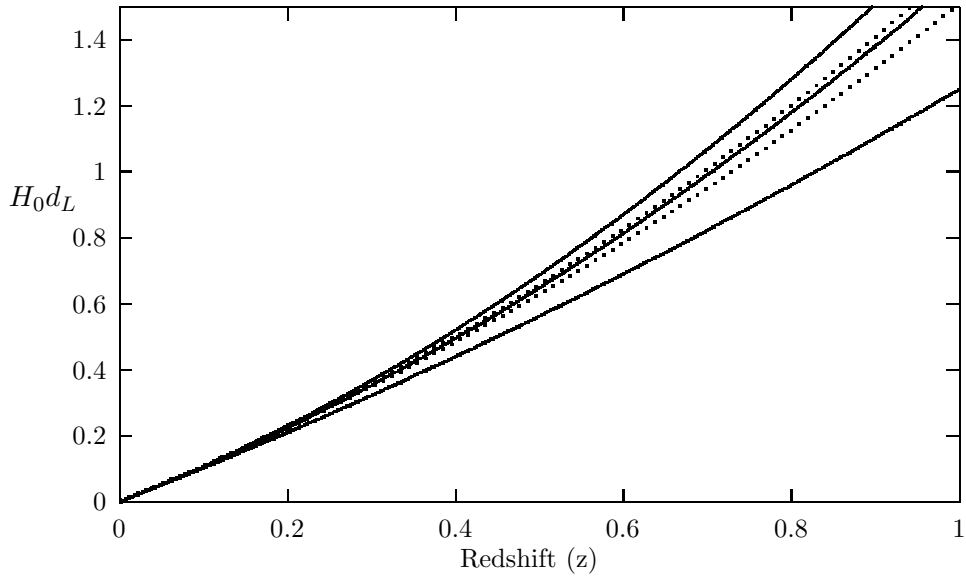

\input missing1.tex
\caption{The luminosity distance versus redshift in different cosmologies.
The lower solid line corresponds to $\Omega_m=1, 
\ \Omega_\Lambda=0$ and $\alpha = -{1\over 2} \ {\rm or} \ -{2\over 3}$, 
Einstein-de Sitter, the middle solid line corresponds 
to $\alpha=-{2\over 3}, \ \Omega_m=0.31, \ \Omega_\Lambda=0.68$, the upper 
solid line corresponds to 
$\alpha=-{2\over 3}, \ \Omega_m=0, \ \Omega_\Lambda=1$ de Sitter, the lower 
 and upper dot lines correspond to $\alpha=-{1\over 2}, \ \Omega_m=0.31, \ 
\Omega_\Lambda=0.68$ and $\alpha=-{1\over 2}, \ \Omega_m=0, \ 
\Omega_\Lambda=1$ de Sitter respectively. The effects of different models 
are observable at high redshift rather than at low redshift.}
\label{fig2.3}
\end{figure}

We can write the luminosity distance relation as
\begin{eqnarray}
H_0 d_L = z + {1 \over 2}(1-q_0)z^2 
\label{f51}\end{eqnarray}
where 
\begin{eqnarray}
q_0 =  {1 \over 2}\Omega_m + ({1+3\alpha \over 2})\Omega_\Lambda ,
\label{f52}\end{eqnarray}
$q_0$ is the deceleration parameter, $\alpha$ is the ratio of pressure to 
energy density of missing energy, and its significance is described in
Chapter 1. The deviation of linear law depends on equation of state 
$\alpha$, the ratio of pressure to energy density of missing energy as 
seen in Fig. \ref{fig2.3}  \\ 

\chapter{Observational Constraints}

\noindent
While developing science and technology, astrophysical objects become more 
visible to us, leading to remarkable developments in cosmology in the   
last few years. The constraints on cosmological parameters have been 
published from the analysis of different observational data. The 
constraints on cosmological parameters differ from one observation to 
another. Sometimes the conflict is very strong. 
We partake in resolving this contradiction by paying rigorous 
attention to statistics, and by combining data from recent observations.
In a sequence of Papers I-III we combined all types of data
having a published error on $\Omega_m$ and $\Omega_{\Lambda}$, using simple
$\chi^2$ analysis. This, of course, implied believing in the errors
and treating them as Gaussian. The techniques and observations usually 
attributed by the astrophysicist for constraining  
cosmological parameters are summarized in this Chapter, and the results
of our $\chi^2$ analysis are discussed. 

\section{Cosmic Microwave Background Radiation}

\noindent
The Cosmic Microwave Background (CMB) is one of the important probes of the 
early Universe, and it was dramatically discovered in 1965 by Penzias and 
Wilson. The first result on temperature variations was published by       
the Cosmic Background Explorer (COBE) satellite 
(Smoot et al. 1992). The analysis was done in a two-parameter space: 
the scalar tilt of density fluctuation of the power spectrum $n_s$, and the 
CMB quadrupole normalization {\it Q}. The power spectrum $P(k)$ can be 
written as $P(k) \propto k^n$, where $n$ is the tilt of primordial spectrum. 
 
The primary aim of the CMB experiments is to determine the power  
spectrum, $C_l$, of the CMB as a function of multipole moment $l$. Given a 
map of the temperature $T({\rm\hat{n}})$ in each direction ${\rm\hat{n}}$ on 
the sky, the power spectrum can be obtained by expanding in spherical 
harmonics, 
\begin{eqnarray}
\left|a_{lm}\right| = \int d{\rm\hat{n}} \ Y_{lm}({\rm\hat{n}}) \ T({\rm\hat{n}}), 
\label{f2.1}\end{eqnarray}
and then squaring and summing the coefficients, 
\begin{eqnarray}
C_l = {1 \over 2l+1} \sum_{m}\left|a_{lm}\right|^2.
\label{f2.2}\end{eqnarray}
If the map covers a patch of the sky that is small enough to be approximated 
as a flat surface, the power spectrum can be written in terms of Fourier 
coefficients: 
\begin{eqnarray}
T_{\vec l} = \int d{\rm\hat{n}} \ e^{-i\vec{l}.\vec{\theta}} \ 
T({\rm\hat{n}}), 
\label{f2.3}\end{eqnarray}
and then 
\begin{eqnarray}
C_l \simeq  \langle \left| T_{\vec l} \right|^2 \rangle,
\label{f2.4}\end{eqnarray}
where the average is taken over all Fourier coefficients $\vec{l}$ that have 
amplitude $l$. Thus, each multipole moment $C_l$ measures the rms 
temperature fluctuation between two points separated by an angle 
$\theta \simeq (l/200)^{-1}$ degrees on the sky. 

For constraining the cosmological parameters, one can fit the data 
in different parameter spaces. Different groups have 
analyzed the COBE DMR data. Lineweaver (1998) and Tegmark (1999) have 
analyzed the CMB data up to $l \simeq 800$. In the Cold Dark Matter (CDM) 
model, Lineweaver (1998) used the parameter space as follows: the 
CDM densities ($\Omega_{cdm}$), baryons ($\Omega_b$), the tilt of scalar 
fluctuation ($n_s$), the CMB quadrupole normalization for scalar 
fluctuation ({\it Q}), the vacuum energy ($\Omega_\Lambda$) and the Hubble 
parameter ({\it h}). 
Tegmark (1999) has added three more parameters in his analysis, tilt of 
tensor fluctuations ($n_t$), the reionization optical depth ($\tau$),
and the CMB quadrupole normalization for tensor fluctuations ({\it r}).  
Moreover, for the best theoretical model, Tegmark and 
Zaldarriaga (2000) have added one more parameter ($\Omega_\nu$) 
for the neutrino mass density parameter in their new analysis.  
So the parameter space is then ten, and the range 
of all parameters can be seen in Tegmark (1999) and Tegmark and 
Zaldarriaga (2000). However, some of the parameters could be dropped out from 
the analysis because their influence was negligible. The  6-dimensional 
likelihood function was then 
integrated over the remaining parameters. Overall, the observations of
CMB are shown by the plot of the multiple moment $C_l$ against the multiple 
{\it l}. Recently, de Bernardis et al. (2000) and Balbi et al. (2000) 
have published the BOOMERANG and MAXIMA-1 data. The angular power spectrum 
from their fitting can be seen in figure \ref{B+Mfig}.  

We use Lineweaver (1998) in our least square analysis. The best fit 
point $O$ is located asymmetrically within a wedge at (0.45, 0.35) 
(Fig. 2 in Lineweaver 1998). Denoting the distance from $O$ to an
arbitrary point $P$ in the plane by $r$, and the distance from $O$
through $P$ to the $1\sigma$ wedge line by $r_0$, our constraint is
therefore of the form $r^2/r_0^2$. In the maximum likelihood we fit CMB  
(Tegmark 1999) likelihood contours with a fifth order polynomial.  
The likelihood contours in the 
($\Omega_m, \Omega_{\Lambda}$) plane available to us from this compilation 
correspond to 68\% ($1\sigma$) and 95\% ($1.96\sigma$) confidence, 
respectively.

\section{Supernovae of Type Ia Observations}

One of the most energetic phenomena observed in the galaxies is the 
explosion of a star into a supernova. The supernovae are divided into two
basic types, Ia and II, and these types are based on the hydrogen spectral 
lines. Type Ia supernovae are brighter than those of type II. Enormous 
luminosities of type Ia can be compared with luminosities of a standard 
candle. The distance-luminosity relationship is used to determine the 
cosmological parameters.

The key approach to determining the cosmological parameters from the 
supernovae observations is in the relation between the apparent magnitude
$(m)$, the absolute magnitude $(M)$, and the luminosity distance $(D_L)$ 
(Riess et al. 1998, Perlmutter et al. 1999):
\begin{eqnarray}
m(z) = M + 5logD_L(z;\Omega_m, \Omega_\Lambda, H_0) +25 .
\label{f2.3}\end{eqnarray}
One can exclude $H_0$ from Eq. (\ref{f2.3}) by writing 
\begin{eqnarray}
m(z)- \mathcal{M} = 5logD_L(z;\Omega_m, \Omega_\Lambda) +25,
\label{f2.4}\end{eqnarray}
where $\mathcal{M}\equiv M - 5logH_0$. 
The luminosity distance $(D_L)$ of SN Ia can be written 
\begin{eqnarray}
D_L = \left( \mathcal{L} \over {4\pi \mathcal{F}}\right)^{1\over 2},
\label{f2.5}\end{eqnarray}
where $\mathcal{L}$ is the intrinsic luminosity and $\mathcal{F}$ is the
observed flux of the supernova. 

In the Friedman-Robertson-Walker cosmologies,
the luminosity distance at a given redshift $z$ is a function of cosmological
parameters. Giving constraints on these parameters the Hubble parameter 
$H_0$, the vacuum energy density $\Omega_\Lambda$, and the mass density  
$\Omega_m$ (Caldwell et al. 1998, 
Garnavich et al. 1998, Riess et al. 1998, Perlmutter et al. 1997, 1999), 
the luminosity distance is
\begin{eqnarray}
D_L = cH_0^{-1}(1+z)\left|\Omega_k\right|^{-1/2} \nonumber \\ 
{\rm sinn} \left \{ \left|\Omega_k\right|^{1/2}  
\int_0^z dz \left[(1+z)^2(1+\Omega_m z)-
z(2+z)\Omega_\Lambda \right]^{-1/2}\right \},     
\label{f2.6}\end{eqnarray}
where $\Omega_k = 1 - \Omega_m - \Omega_\Lambda$, sinn is sinh for positive
curvature ($\Omega_k \geq 0$) and sinn is sin for negative curvature 
($\Omega_k \leq 0$). 

The distance moduli $f(H_0, \Omega_m, \Omega_\Lambda)$ can be obtained from  
the probability density function (pdf) of these parameters.  
The likelihood for the cosmological parameters can be 
determined from the $\chi^2$ statistics, where $\chi^2 \sim 
f(H_0, \Omega_m, \Omega_\Lambda$). The $\Omega_m$ and $\Omega_\Lambda$ are 
independent of the Hubble parameter $H_0$. Therefore, one can convert the  
three-dimensional pdf to a two-dimensional one as follows
\begin{eqnarray}
pdf(\Omega_m, \Omega_\Lambda) = \int_{0}^{+\infty}
pdf(\Omega_m, \Omega_\Lambda, H_0)dH_0.  
\label{f2.7}\end{eqnarray} 

High-Z Supernovae Search Team (HSST) and Supernovae Cosmology Project
(SCP) have searched supernovae at different redshift. HSST (Riess et al.
1998) has discovered 16 SNe Ia in the redshift range 0.16 - 0.62 and 34 nearby
supernovae for the constraining on $H_0$, $\Omega_m$, $\Omega_\Lambda$,
the deceleration parameter $(q_0)$, and the dynamical age of the Universe
$(t_0$), whereas the Supernova Cosmology Project (Perlmutter et al. 1999) 
has discovered 42 supernovae in the redshift range 0.18 - 0.83 for the 
constraining on cosmological parameters.  

The HSST team has used two different methods, a template method and 
Multicolor Light Curve Shape (MLCS) method respectively, for the data 
fitting. The differences of the two methods can be distinguished from 
the contour plot in their fits (their Figs. 6 and 7).  
A large portion of the contour plot for the MLCS method is  
in the unphysical region $\Omega_m$ $<$ 0. This can be improved by 
using Feldman and 
Cousins (1998) statistics. This is a classical statistics, having the 
advantage that is always pulls the fit from the unphysical region to 
a physical region.
However, this could only be done when the original data were analyzed.
Their best value is in the $(\Omega_m, \Omega_\Lambda)$ plane (0.20,0.65). 
The likelihood contours in the ($\Omega_m, \Omega_\Lambda$) 
plane available to us from this compilation correspond to 68.3\% ($1\sigma$), 
95.4\% ($2\sigma$), 99.7\% ($3\sigma$) confidence, respectively. We have  
used this data set for both the $\chi^2$ and the maximum likelihood analyses. 
In our fit these 
constraints are represented by a term in the $\chi^2$-sum of the form 
\begin{eqnarray}
[w^2 + z^2 -(\sigma^2_w -\sigma^2_z) (w^2/\sigma^2_w)]/\sigma^2_z\ ,
\label{f3.8}\end{eqnarray}
where $w$ and $z$ are the rotated coordinates
\begin{eqnarray}
w &= (\Omega_m - \Omega_{m,0}) \cos\theta + (\Omega_{\Lambda} -
\Omega_{\Lambda,0})\sin\theta\nonumber\\
z &= -(\Omega_m - \Omega_{m,0}) \sin\theta + (\Omega_{\Lambda} -
\Omega_{\Lambda,0})\cos\theta \ .
\label{f3.9}\end{eqnarray}
The rotation angle is $\theta = 51^{\circ}.1$ and the $w$ and $z$
errors are $\sigma_w = 1.27, \sigma_z = 0.18$ for the HSST, whereas for the 
SCP, the rotation angle is $\theta = 54^{\circ}.35$ and the $w$ and $z$
errors are $\sigma_w = 1.3, \sigma_z = 0.15$. The same data set has been  
analyzed with maximum likelihood, and approximating the published confidence 
contour by a fifth order polynomial.  

The sample of the Supernova Cosmology Project is larger than HSST. 
Due to the large sample, 
the statistical uncertainty is small and the confidence region is narrow.
It is also possible to estimate some systematic uncertainties in this case.
In the flat case the value found is $\Omega_m = 0.28 \pm 0.085 ({\rm stat})
\pm 0.05 ({\rm syst})$. The two-dimensional contour plot   
in the parameter space ($\Omega_m, \ \Omega_\Lambda$) plane can be seen 
in fig. \ref{snfig}. We have used this data in our both analyses.
None of the teams has considered the effect of galactic dust on the light 
curve. The effect of intergalactic dust on light curve has been described 
by Aguirre (1999). 
However, both supernova projects have concluded that the expansion of the 
Universe is accelerating. The coming SNAP satellite will be able to measure an
enormous amount of distant supernovae every year. This can   
provide us essential information on the critical nature of the Universe.

\section{Classical Double Radio Galaxies}

Powerful extended classical double radio galaxies are used for 
constraining global cosmological parameters. It is much like the
relation between supernovae and standard candles, comparing  
coordinate distance and powerful radio galaxies. At first Daly (1994, 1995) 
applied this method for constraining cosmological parameters and 
later adopted by others  
Guerra \& Daly (1996, 1998), Guerra, Daly \& Wan (2000).

It is observed that at a given redshift all radio sources
have a similar size. This size may be estimated in two different
ways: i) by the average size of the full population of powerful extended
radio galaxies at that redshift; ii) by the product of average rate of 
growth of the source and the total time for the production of powerful jets.  
The two measures depend on the angular size distance
to the sources, and the ratio of the two measures depend on the cosmological
parameters $\Omega_m, \Omega_\Lambda$, and a model parameter $\beta$.
The rate of growth of the radio sources strongly depends on the redshift:   
it increases with redshift, but the average size of the full population
decreases monotonically with redshift, for a redshift greater than 0.5. 

In 1998, Daly, Guerra and Wan studied fourteen samples for constraining 
cosmological parameters.  
Guerra, Daly and Wan (2000) have identified 70 classical double radio 
galaxies with a redshift between zero and two. From there they have studied 
20 of average size for constraining cosmological parameters.  
However, the likelihood contours in the ($\Omega_m, \Omega_{\Lambda}$) 
plane available to us from this compilation correspond to 68\% ($1\sigma$) 
and 90\% ($1.64\sigma$) confidence, respectively. These confidence regions
are quite large. This is therefore one of the weak constraints, 
thus affecting very little the total fit. For the $\chi^2$ analysis  
we introduced this constraint as a mathematical expression  
like Eq. (\ref{f3.8}) and Eq. (\ref{f3.9}), where the rotation angle is 
$\theta = 70^{\circ}.4$ and  the $w$ and $z$
errors are $\sigma_w = 0.84, \sigma_z = 0.33$. The same data have been 
introduced in the maximum likelihood analysis as a fifth order polynomial.

\section{Gravitational Lensing}
According to Einstein's Theory of Relativity 
light rays are deflected due to the gravity of massive bodies. The deflection 
of light by massive bodies is known as Gravitational 
Lensing. From the gravitational lensing statistics, the  matter 
distribution in the 
Universe can be studied. Distant objects (i.e. QSO's) are usually used 
as a source of light. Galaxies or clusters of Galaxies are used as a 
lensing plane and on the Earth we are the observers. 
From this geometry one can establish a relation between an 
angular distance and the cosmological parameters at a given
redshift.

Different techniques have been applied for constraining the cosmological
parameters. Using the results from 5 optical quasar surveys, Cheng and 
Krauss (1998) have re-analyzed lensing statistics, and the best fit of  
$\Omega_m$ gives the maximum likelihood in the 
range between 0.25-0.55 in the flat Universe. Systematic uncertainties is  
dominant in this analysis, and the uncertainty comes mainly from the galaxy 
luminosity function and dark matter velocity dispersion. From the redshift
survey of radio and optical data, Falco et al. (1998) have fitted
the data independently and combined them. The confidence level is
so wide that we can not even reach the lower limit of the contour.

Chiba \& Yoshii (1999) have presented new calculations of gravitational
lens statistics in view of the recently revised knowledge of the
luminosity functions of elliptical (E) and lenticular (S0) galaxies and their
internal dynamics. They applied their revised lens model to a sample of
867 unduplicated QSOs at $z > 1$ taken from several optical lens
surveys, as well as to 10 radio lenses. In sharp contrast to the previous
models of lensing statistics that have supported a high-density universe
with  $\Omega_m=1$, they concluded that a flat universe with 
$\Omega_m=0.3^{+0.2}_{-0.1}$ casts the best case to explain the results
of the observed lens surveys.
Instead of using the above quoted $\Omega_m$ value, we use the 68\%
likelihood contour in the two-dimensional parameter space of Fig. 8 of
Chiba \& Yoshii (1999). We integrate out the characteristic velocity 
dispersion ($\sigma^*$), and we thus obtain
the one-dimensional 68\% confidence range $\Omega_\Lambda = 0.70 \pm 0.16$. 

A preliminary result from the Cosmic Lens All-Sky 
Survey (CLASS) has been presented by Helbig (2000). Most probably this is 
the strongest constraint on the $(\Omega_m, \ \Omega_\Lambda)$ 
plane from lensing surveys. Although the maximum 
likelihood is wide, this is a strong constraint on $\Omega_\Lambda$. A 
large portion of the confidence level is in the region $\Omega_\Lambda$ $<$ 0 
which may be considered unphysical. The upper bound of the contour may be 
interesting for $\Omega_\Lambda$ from this compilation. At 
this moment we get very little benefit from this result. However, we expect 
that in the near future it would be able to provide us with a strong 
constraint on the cosmological parameters. The likelihood contours in the 
($\Omega_m, \Omega_\Lambda$) plane available to us from this 
survey correspond to 68.3\% ($1\sigma$), 90\% ($1.64\sigma$), 
95\% ($1.96\sigma$), 99\% ($2.58\sigma$) confidence, respectively.

\section{Cluster Mass Function and Ly$\alpha$ Forest}

The most common assumption is that the structure we observe today such 
as galaxies, clusters of galaxies and voids results from the growth of 
primordial density fluctuations by gravitational instability. These 
fluctuations are normally assumed to have originated from a Gaussian random 
process. The overdense regions will decelerate faster than the background 
Universe, because of their large gravitational field, resulting in an 
increase of their contrast relative to the background. If this deceleration 
is large enough, these regions will turn back and recollapse on themselves, 
resulting in the formation of positive density structures such as galaxies 
and clusters. The opposite phenomenon occurs in underdense regions. These 
regions decelerate more slowly than the background Universe, thus getting 
more underdense, and eventually become the cosmic voids we observe today. 

Croft et al. (1998) have presented a theoretical model for recovering the
linear power spectrum of mass fluctuations from high-redshift 
Quasi-stellar Object (QSO) Ly$\alpha$ spectra. They have tested the model
by using 16 QSOs. For the precise measurement of $P(k)$ Croft et al. (1999) 
have combined 3 more samples at redshift 2.5 with their previous 
determination. 
Most recently, toward a more precise measurement of matter 
density, Croft et al. (2000) have used a wide number of samples (53 QSOs)      
of Ly$\alpha$ forest spectra in their analysis,   
discovered by the Keck telescope at redshifts between 2 and 4. 
Weinberg et al. (1999) estimated  $\Omega_m$ by combining the cluster
mass function constraint with the linear mass power spectrum  determined
from the Ly$\alpha$ data (Croft et al. 1998, 1999). 
For $\Omega_{\Lambda}=0 $ they obtained
$\Omega_m=0.46^{+ 0.12}_{-0.10}$ and for a flat universe they obtained
$\Omega_m=0.34^{+ 0.13}_{-0.09}$. 
In the flat cosmology Croft et al. (2000) obtained  
$\Omega_m=0.50^{+ 0.13}_{-0.10}$.

\section{Gas fraction in X-ray clusters}

Clusters emit X-rays which indicates that the clusters consist of
a large amount of hot gas. From the measurement of the gas fraction in X-ray
clusters one can reach the Universal matter density $\Omega_m$. It is
observed that at a certain radius,  
the clusters are in hydrostatic equilibrium. This radius is 
known as the {\it virial radius}. A common definition of the virial
radius is $R_{500}$; outside this radius the density drops below
500 in units of the critical density (Navarro, Frenk \& White 1995). 
All matter outside the virial surface are infalling with the cosmic 
mix of components. Then the baryonic mass fraction measured at the virial
radius must be unbiased. Thus, by measuring the gas
fraction near the virial radius, one expects to obtain fairly unbiased
information on the ratio of $\Omega_m$ to the cosmic baryonic density
parameter $\Omega_b$. This type of analysis has been done before 
(Evrard 1997) using the best known value of $\Omega_b$ to derive 
a value of $\Omega_m$. In Chapter 5, we do the opposite: we use 
our best estimation of $\Omega_m$ to determine some poorly known parameters. 

For this purpose Evrard (1997) has used a very large sample of clusters:
the ROSAT compilation of David, Jones \& Forman (1995) and the Einstein
compilation of White \& Fabian (1995). He has obtained a realistic value of 
\begin{eqnarray}
{\Omega_m\over \Omega_b} h^{-4/3} \approx (11.8\pm 0.7)\ .
\label{f2.8}\end{eqnarray}
This value includes a galaxy mass estimate of 20\% of gas mass, and a
baryon diminution $\Upsilon(500)=0.85$ at $R_{500}$ (a detail 
definition of $\Upsilon$ can be found in Section 5.4). To get a precise 
information on the density parameter $\Omega_m$, we need the precise values 
of $\Omega_b$ and $h$. The problem lies in the contradiction in baryonic 
density between the Big Bang Nucleosynthesis (BBN) and the Cosmic Microwave 
Background (CMB) observations by the recent balloon experiments 
BOOMERANG (2000) and MAXIMA-1 (2000). However, 
taking $\Omega_b=0.020\pm 0.001 h^{-2}$ {\rm (68\% CL)} from the low primordial
deuterium abundance (Burles et al. 2001), and $h=0.73\pm 0.07$ {\rm (68\% CL)}
Gibson \& Brook (2001) who have re-analyzed Calan-Tololo and the Center for 
Astrophysics (CfA) SNe Ia datasets, one obtains $\Omega_m=0.29\pm 0.11$.

\section{Power Spectrum of Galaxy Fluctuations}

Matter in every direction appears to be distributed in high-density
peaks separated by voids. The average separation distance is 
$\sim130h^{-1}$Mpc, which translates into a peak in the power spectrum 
of mass fluctuations. This provides a co-moving scale for measuring 
cosmological curvatures. Broadhurst \& Jaffe (1999) used a set of 
Lyman galaxies at $z\sim 3$ finding a constraint in the form
$\Omega_m =  0.20\pm 0.10 + 0.34 \Omega_{\Lambda}$. 
Roukema \& Mamon (1999) have carried out a similar analysis of quasars, 
finding $\Omega_m =  0.24\pm 0.15 + (0.10\pm 0.08)\Omega_{\Lambda}$.

The distribution of galaxies is considered to be Gaussian. For the peaks 
measurements, small angle geometry with the normal
could provide precise information of the co-moving scale. The 
density parameters $\Omega_m$, $\Omega_\Lambda$ measured from these  
observations lie between the locii
of supernovae and cosmic microwave background, that is, 
$(3\Omega_m - \Omega_\Lambda)$.

\section{Galaxy Peculiar Velocities}

The large-scale peculiar velocities of galaxies correspond via gravity to
mass density fluctuations about the mean, and depend also on the mean
density itself. Two catalogs of galaxies have been analyzed for these
velocities in order to provide information on $\Omega_m$: the Mark III
catalog (Willick et al. 1997) of about 3000 galaxies within a distance of
$\sim 70 h^{-1}$ Mpc, and the SFI catalog (Borgani et al. 1999) of about
1300 spiral galaxies in a similar volume. Combining the results in these
catalogs, Zehavi \& Dekel (1999) quote the constraint
\begin{eqnarray}
\Omega_m \ h_{65}^{1.3} \ n^2\simeq 0.58\pm 0.12\ ,
\label{f2.9}\end{eqnarray}
in the case of flat cosmology,
where the error corresponds to a 90\% confidence level. Taking the index
$n$ of the mass-density fluctuation power spectrum to be $n=1.0\pm 0.1$
(Bond \& Jaffe 1998), one obtains the constraint
$\Omega_m = 0.55 \pm 0.14$,
where the error corresponds to a 68\% confidence level.

\section{X-ray Cluster Evolution}
The Clusters of Galaxies are the largest known gravitationally bound
and the most massive objects in the Universe. Massive clusters are 
more luminous and they emit X-rays. It is observed that the evolution 
of the number density of rich cluster galaxies breaks the degeneracy 
between the Universal mass density and the normalization of the  
power spectrum $\sigma_8$. This evolution is   
strong with high-mass and low-$\sigma_8$ models. The number 
density of clusters decreases by a factor of $10^3$ from $z=0$ to $z=0.5$. 
The same clusters show mild evolution in low-mass and high-$\sigma_8$ 
models, the decrease is by a factor of 10. Bahcall et al. 
(1997, 1998), Eke et al. (1998), Viana \& Liddle (1999), Moscardini et al.
(2000) have used this degeneracy as a powerful tool for the 
constraint on cosmological parameters. Note that this evolution is 
strongly model dependent. In the next Chapter we have analyzed 
the recent Advanced Satellite for Cosmology and Astrophysics (ASCA) data 
for the study of baryonic dark matter.

There are still a few more constraints that have been used in the 
$\chi^2$ analysis. 
For critical evaluation of the density parameters $\Omega_m$ and 
$\Omega_\Lambda$  we have done another combined study in the Paper (IV). The 
rest of the constraints have been discussed in Paper (IV), and they will be 
discussed also in the next Chapter. The data used in the $\chi^2$ method 
are summarized in tables \ref{table1} and \ref{table2}.


\begin{table}[h]
\begin{center} 
\begin{tabular}{lcccc}  
\hline
{\em Sources}  &{\em $\Omega_m$} &{\em $\Omega_\Lambda$} & 
{\em $\Omega_0=\Omega_m+\Omega_\Lambda$} &{\em References} \\ \hline
High-Z SN Ia    &  $0.20\pm{0.22} $   & $0.65\pm{0.28}$     & $0.85\pm{0.35} $ & \cite{ries}    \\
SCP    & $0.28_{-0.13}^{+0.19}$  & $0.73_{-0.22}^{+0.19}$  & $1.01_{-0.25}^{+0.26}$  & \cite{per1}   \\
Double Radio Galaxies   &  $0.09_{-0.09}^{+0.29}$  &  correlated  & -- & 
\cite{guer}   \\
Various CMBR data   &  correlated   & correlated   & --  & \cite{lin1}  \\ 
BOOMERANG CMBR  & --  & --  & $1.03_{-0.06}^{+0.07}$ & \cite{debe}  \\
MAXIMA-1 CMBR  & --  & --  & $0.90 \pm{0.08}$ & \cite{balb}  \\
\hline  
\end{tabular}
\end{center}
\caption {Two-dimensional data in the $\chi^2$ analysis. All 
errors are given at 1$\sigma$ confidence level.}
\label{table1}
\end{table}


\begin{table}
\begin{center} 
\begin{tabular}{lccc}  
\hline
{\em Sources}  &{\em $\Omega_{m}^{flat}$}&{\em $\Omega_{\Lambda}^{flat}$}  &{\em References} \\ \hline
Galaxy Power Spectrum & $0.40\pm{0.10}$ &$1-\Omega_m$    & \cite{broa}  \\
Gravitational Lensing & $1-\Omega_\Lambda$ &$0.70\pm{0.16}$ & \cite{chib} 
  \\
Gravitational Lensing &$1-\Omega_\Lambda$ &$0.64\pm{0.15} $ & \cite{immg}
   \\  
Ly$\alpha$ Forest  & $0.34_{-0.09}^{+0.13}$ &$1-\Omega_m$  & \cite{wei1}
 \\  
Temperature-Redshift & $0.27\pm{0.10} $    & $1-\Omega_m$ & \cite{dona} 
   \\  
Galaxy Peculiar Velocities & $0.55\pm{0.14} $   &$1-\Omega_m$ & \cite{zeh1}  \\
X-ray Clusters &  $0.36\pm{0.09} $    & $1-\Omega_m$ & \cite{evr1} \\ 
Large Scale Structure &$0.30\pm{0.15} $ &$1-\Omega_m$ & \cite{rouk}  \\  
Cluster Evolution & $0.45\pm{0.20} $ &$1-\Omega_m$ &  \cite{eke1} \\

\hline  
\end{tabular}
\end{center}
\caption {One-dimensional (flat space) data in the $\chi^2$ analysis. 
Errors are given at 1$\sigma$ confidence level.}
\label{table2}
\end{table}

\section{Data Analysis: $\chi^2$ Method}

The $\chi^2$ statistics is one of the strongest mathematical tools that has  
been successfully used for experimental and observational data analyses.  
It is used to investigate whether the distributions differ from one 
another, and of course to find the best estimation from a set of data. A 
confidence level is also determined from the combined distribution. 
The $\chi^2$ statistics is calculated as 
\begin{eqnarray}
\chi^2 = \sum_{i} {(O_{i} - E_{i})^2 \over e_{i}},
\label{f2.10}\end{eqnarray}
where $O_i^{'s}$ are the observed and $E_i^{'s}$ are the expected values. 

Using two free parameters 
$\Omega_m$, $\Omega_\Lambda$, we have performed 
a $\chi^2$ fit to the constraints summarized in the
tables \ref{table1} and \ref{table2}, assuming that all reported 
observational errors as well as systematic errors are Gaussian. 
For this analysis we use MINUIT standard program 
for function minimization and error calculation (James \& Roos), which is 
available from  the CERN Program Library and is documented there 
as entry D506. We fit the data in two different 
cases: one-dimensional and two-dimensional. 
The two-dimensional best fit value is found to be  
$\Omega_m = 0.33 \pm 0.07$, $\Omega_\Lambda = 0.66 \pm 0.12$,  thus
$\Omega_m + \Omega_\Lambda = 0.99 \pm 0.14$, where the errors are  
estimated for 1$\sigma$ {\rm (CL)}. In the exact flatness case we have 
found $\Omega_m = 0.33 \pm 0.04$, $\Omega_\Lambda = 0.67 \pm 0.04$.
The interesting thing is that there is no remarkable change in the 
parameters but the errors decrease because the dimensional space is one. 
To find the absolute 1$\sigma$ and 2$\sigma$ confidence   
regions in the $\Omega_m$, $\Omega_\Lambda$-plane we add 
$\Delta \chi^2$ = 2.3 and 6.2 up from the minimum, respectively. 
Our $\chi^2$ fit is extremely good. That can 
be seen from the degrees of freedom. In the two-dimensional 
fit and for ten constraints, $\chi^2$ is 4.42, where the degrees of 
freedom is 8, and in the one-dimensional case $\chi^2$ is 4.43 where the 
degrees of freedom is 9. In the two-dimensional case, a fit is good when 
$\chi^2$ is about N - 2, where N is the total number of constraints.   
Therefore the goodness-of-fit here is extremely high. 

It is obvious that the greater the discrepancy between observed and 
adjusted values, the greater will be the value of $\chi^2$. To check
this argument we have measured the individual $\chi^2$ for each 
observational data set. In our ten constraint analysis the maximum 
$\chi^2$ comes from the CMBR data, but it is still very small: slightly 
less than one.  

We have also fitted the data without the SN Ia constraints.
We found very little change. The result is then 
$\Omega_m = 0.34 \pm 0.07$, $\Omega_\Lambda = 0.64 \pm 0.09$,  thus
$\Omega_m + \Omega_\Lambda = 0.98 \pm 0.11$. The reason is that there are
strong constraints, for instance, gravitational lensing on   
$\Omega_\Lambda$  and gas fraction in X-ray clusters on the $\Omega_m$, 
and they constrain the result similarly. We did not measure any 
systematic error from this fit. If we carefully observe, we do not
need to estimate the systematic errors. The data we have used here pull in 
randomly different directions. It means that the CMB constraint 
is laying along the flat line, supernova constraints are orthogonal 
to the CMB constraint. Again the gas fraction in X-ray 
cluster constraints is perpendicular to the gravitational lensing, and the  
distribution of others is quite random. Thus, the total effect might be a  
 mutual cancellation. So we can say that the systematic errors are already
included. In such situation there is no reason to reconsider the  
effects of systematic errors. 

There can be more arguments against the systematic errors. Why does one
neglects them in this analysis? In fact, most of the constraints 
have come from fits of higher dimensions. We do not have any information 
whether the parameters have been properly rescaled or not for the lower 
dimensionality. Overall, 
we think that our errors have already been generous, and there is no 
motivation to add further systematic errors arbitrarily. 

Later, we combined a few more constraints in our $\chi^2$ fit Paper (III), 
where the total number of constraints were sixteen. From the overall 
analysis, it was shown that the flatness of the Universe is robust. 

However, it is observed that a large number of data are not Gaussian, 
the best value is also being unknown in some cases. In such a 
situation, we loose information about the input data. Taking these points 
into consideration, a critical analysis has been done in the Paper IV.   
The limitations and advantages of $\chi^2$ and maximum 
likelihood method have been described in Paper IV. Some of the constraints  
that will be used in the maximum likelihood analysis have already been 
discussed in this Chapter.

\vspace{1.5cm}

\begin{figure}[h]
\centering
\mbox{\epsfxsize=11.0truecm\epsfysize=11.0truecm\epsffile{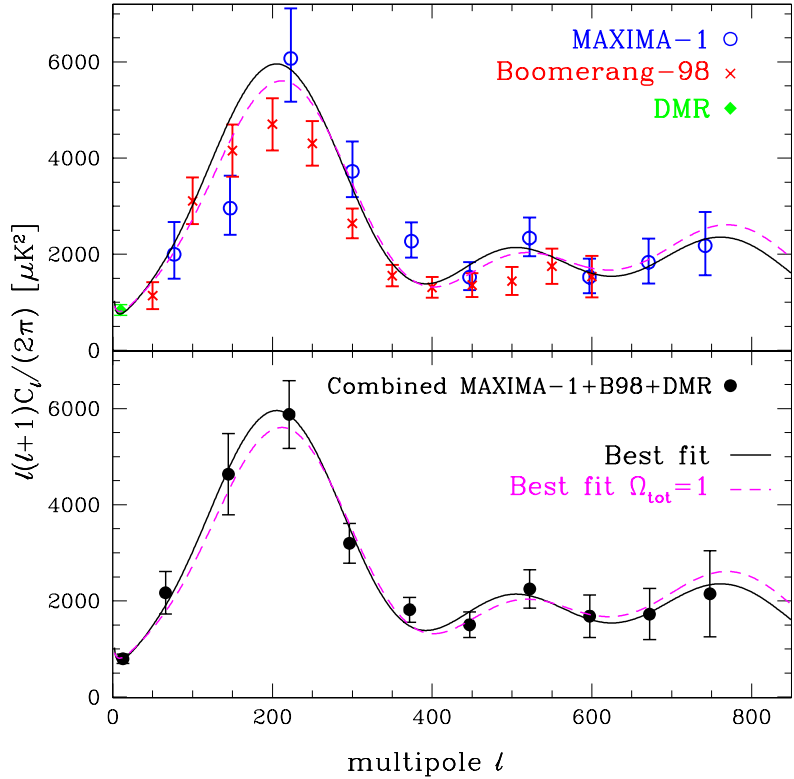}}
\caption{Top: CMB power spectra from BOOMERANG, MAXIMA-1 and COBE-DMR. 
Bottom: Beam calibration uncertainties are used to adjust the peaks 
(Jaffe et al 2000.)} 
\label{B+Mfig}
\end{figure}

\begin{figure}[h]
\centering
\epsfig{file=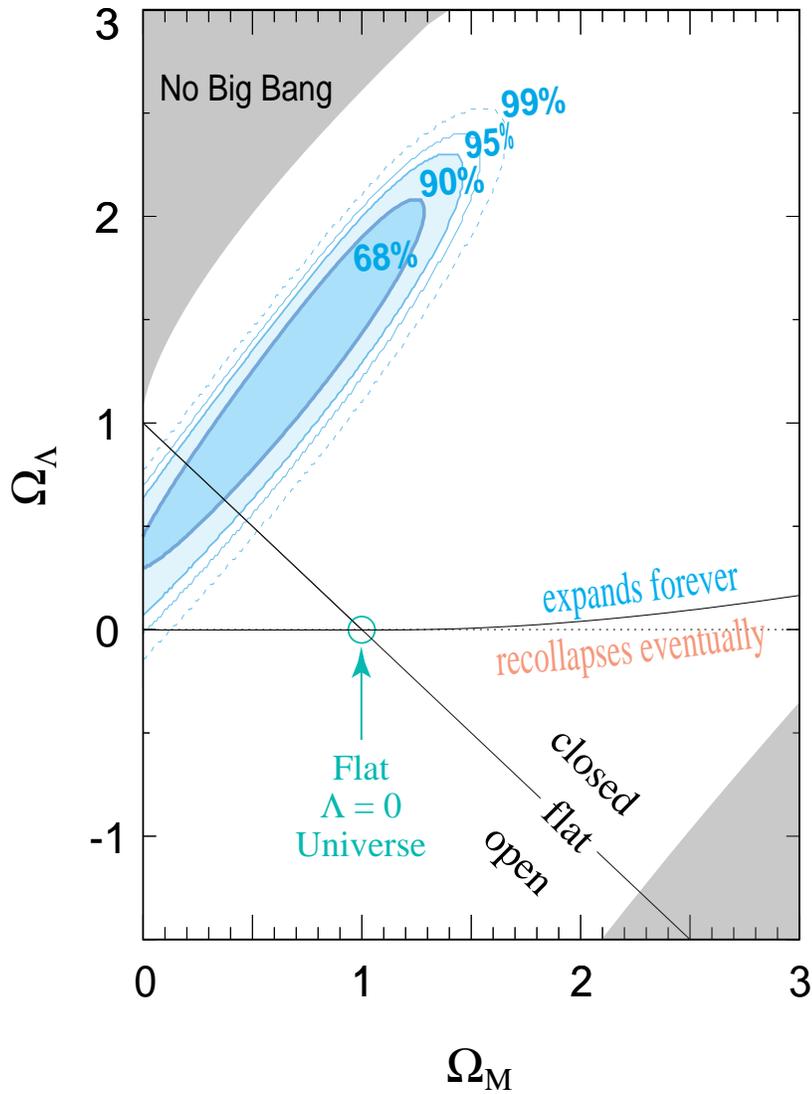,width=11cm}
\caption{The best fit confidence regions in the 
($\Omega_m, \ \Omega_\Lambda$) plane obtained from the analysis 
of the Supernovae Cosmology Project (Perlmutter et al.
1999). The 68\%, 90\%, 95\% and 99\% confidence region are shown. 
The diagonal line corresponds to flat cosmology. Different
cosmologies are also indicated.} 
\label{snfig}
\end{figure}

\chapter{Observational Data Analysis}

Different groups have combined data from 
different observations, e.g. Lineweaver (1999) combined the supernovae 
data with Cosmic Microwave Background (CMB) data, 
X-ray cluster data, cluster evolution
data and double radio sources. Le Dour et al. (2000) analysed
only Cosmic Microwave Background data, whereas Tegmark et al. (2000a) 
combined them with Infrared Astronomical Satellite (IRAS) Large Scale 
Structure (LSS) data. Tegmark \& Zaldarriaga (2000b, 2000c) and Hu
et al. (2000) combined BOOMERANG and MAXIMA data, Melchiorri et al.
(2000) combined BOOMERANG and Cosmic Background Explorer (COBE) data.
Bridle et al. (2000) combined the
Cosmic Microwave Background data with galaxy peculiar velocities and the 
supernovae data. The balloon data have been combined with LSS and 
supernovae data by Jaffe et al. (2000) and by Bond et al. (2000), and 
with a different set of LSS data by Novosyadlyj et al. (2000a) and
Durrer \& Novosyadlyj (2000). We have done a critical combined  
study using Maximum Likelihood method. The main interest herein is the
following: i) taking care of statistics; ii) a combined two-dimensional  
output contour  
and the best value using individual two-dimensional contours as input;    
iii) agreement and disagreement of data with each other. In this analysis 
we have used supernovae data (Riess et al. 1998, Perlmutter et al. 1999),  
CMB data (Tegmark 1999, de Bernardis et. al. 2000, Balbi et al. 2000, 
Hanany et al. 2000), LSS data (Novosyadlyj et al. 2000b) and Double Radio 
Galaxies data (Guerra el al. 2000).

\section{Maximum Likelihood Method}

This is another mathematical tool that has been used for the data analysis. 
From the statistical and theoretical point of view, the best method for the 
data analysis with a high resolution is so far the {\it maximum likelihood 
method}. The likelihood function for a set of data is the joint probability 
density function (pdf) of the data, given some parametric model. The values 
of the parameters maximize the sample likelihood. 
In principle, we can obtain the $\chi^2$ from the maximum likelihood. 
For a normally distributed variable $\chi^2$ is the negative logarithm of 
the maximum likelihood. So, we can say 
that maximum likelihood is optimal, and $\chi^2$ is a special 
case of maximum likelihood. The advantage of the maximum likelihood  
method is that we can use all kinds of data as an input 
with more precision, details of which may be found in Paper (IV).

We have represented the confidence contours of the data by fifth order 
polynomials of the form
\begin{eqnarray}
P(\Omega_m \Omega_{\Lambda}) = \Omega_m^m \Omega_{\Lambda}^n, (m+n)\leq 5 .
\label{f4.1}\end{eqnarray}
There are 20 terms in the polynomials. Therefore we read off 20
points from the $1\sigma,\ 1.64\sigma,\ 2\sigma,\ 3\sigma$
contours and the best value, if available. Here we get a 20 x 20 matrix. 
We invert the matrix once and multiply with the data points that are 
taken from the contours. The equation is then ready to describe the  
nature of data. Since we already
know the approximate location of the globally favored region from
all previous studies, we can take care of that our polynomial
approximation is good over that region. This fit region is
defined by $0.15 \le \Omega_m \le 0.50$ and $0.40 \le \Omega_{\Lambda} 
\le 0.88$, but the sample points are taken also from outside this region 
in order to obtain a well-behaved polynomial inside the region. Far
away from it, the polynomial approximation of course breaks down
completely.

We have taken special care of each data set. We have checked that 
the polynomial is non-negative in the fit region. We have also examined 
the location of the 0.33$\sigma$ contour, in order to verify that the 
region is reasonably centrally located.
Note that one can use higher order polynomials for better accuracy but it
would be very difficult to control the polynomial; even in the fifth order 
polynomial we need some constraints to adjust the data points up to the sixth  
decimal.

In the previous Chapter, we have described the two well known analyses 
from the supernovae observations and their results. Since we know that the
supernovae observations constrain essentially only the  
$\Omega_{\Lambda} - \Omega_m$ parameter combination, let us compare 
their results along the $\Omega_{\Lambda} - \Omega_m$ direction, 
where they have used the same method. Along the flat line the result 
is then $\Omega_{\Lambda} - \Omega_m = 0.44 \pm 0.10$ for 
Supernovae Cosmology Project (SCP) (Perlmutter et al. 1999), 
and $\Omega_{\Lambda} - \Omega_m = 0.36 \pm 0.10$ for the High Redshift 
Supernovae Search Team (HSST) (Riess et al. 1998). 
A 5\% systematic uncertainty have been included in the SCP result but we do 
not have any information about the systematic uncertainty for the HSST case. 
However, the two observations then agree within their statistical errors. 

\begin{figure}[h]
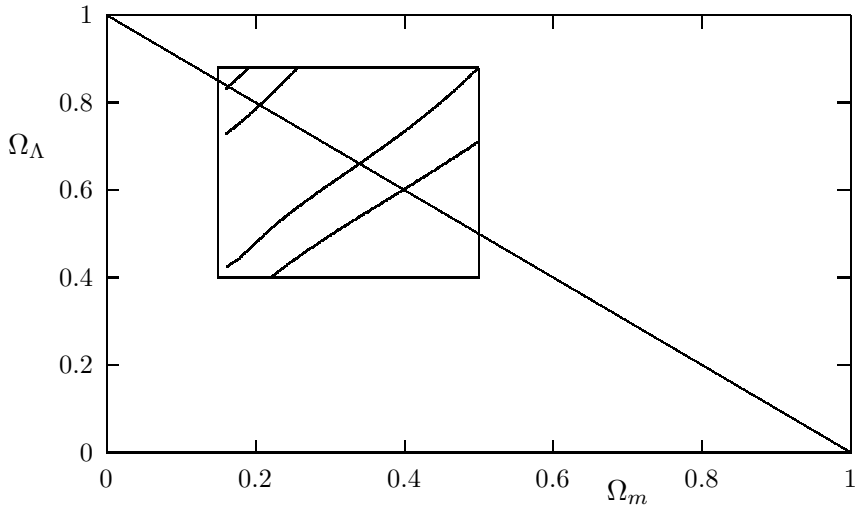

\centering
\input SNsig12.tex
\caption{Confidence contours of the log-likelihood sum of two 
SN Ia observations (HSST and SCP). The inner and outer curves corresponds 
to 1$\sigma$ and 2$\sigma$ in the ($\Omega_m, \ \Omega_{\Lambda}$)-plane,  
respectively. The significance of the square is described in the text.
The diagonal line corresponds to a flat cosmology.}
\label{fig4.1}
\end{figure}

Now we shall see the combined results of these two observations.  
In Fig. \ref{fig4.1}  we show the confidence contours of the log-likelihood
sum of the two observations in our polynomial approximation, drawn
only in the above mentioned ranges of $\Omega_m$ and $\Omega_{\Lambda}$. 
Along the flat line these experiments determine 
$\Omega_{\Lambda} - \Omega_m = 0.45 \pm 0.13$. The SCP result is stronger 
than that of HSST. Therefore, the best fit is moved toward the SCP but 
the combined contour is naturally wider. 

\begin{figure}[h]
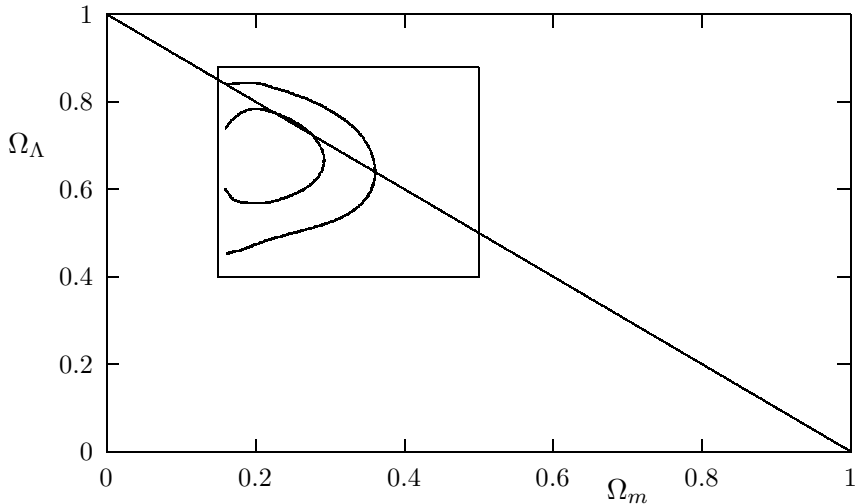

\centering
\input 1to4.tex
\caption{The confidence contours of the log-likelihood sum of two SN Ia 
(HSST and SCP), CMB   
and Double Radio Galaxies. The inner and outer curves correspond to 
1$\sigma$ and 2$\sigma$ in the ($\Omega_m, \ \Omega_{\Lambda}$)-plane, 
respectively. The significance of the square is described in the text.
The diagonal line corresponds to a flat cosmology.}
\label{fig4.2}
\end{figure}

Constraints on cosmological parameters from earlier Cosmic Microwave    
Background Radiation (CMB) and Double Radio Galaxies has also been described 
in the previous Chapter. By CMB we mean only the observations summarized by 
Lineweaver (1998) and Tegmark (1999), not including BOOMERANG and MAXIMA-1 
which we treat separately. CMB puts a strong constraint along the direction   
$\Omega_0 = \Omega_m + \Omega_\Lambda = 1$. The combination of 
supernovae with CMB is also well known. It can be interesting to combine 
Double Radio Galaxies with supernovae and CMB.  
In Fig. \ref{fig4.2}  we show the confidence contours of the log-likelihood
sum of two supernovae (HSST and SCP), CMB and Double Radio Galaxies in our 
polynomial approximation, drawn
only in the ranges of $\Omega_m$ and $\Omega_{\Lambda}$ that we sample. 
The Double Radio Galaxies is so far a weak constraint on 
$\Omega_m$ and $\Omega_{\Lambda}$. Thus the confidence region is mostly  
dominated by supernovae and CMB. 

\section{Balloon Experiments}
The recent two balloon experiments BOOMERANG (de Bernardis et al. 2000) 
and MAXIMA-1 (Balbi et al. 2000, Hanany et al. 2000) have given us a 
most exciting information in observational cosmology. The two 
experiments are held in two different parts of the world. 
BOOMERANG is held in Antarctica and the MAXIMA-1 in California. 
There is a big exposed time difference between two observations. 
BOOMERANG was flown 11 days for the data collection whereas MAXIMA-1 
was flown only for 2 hours. According to their best estimation,  
BOOMERANG has hinted that the Universe is might even be closed. However, 
MAXIMA-1 tells us that the Universe is open. 
On one hand we can not escape from the statistical fluctuations, and 
on the other hand there could be some important corrections in the theory. 
Let us now turn to the theory. The widely used formula is that the  
first peak corresponds to a flat cosmology, and it was 
derived by Kamionkowski et al. (1994) for the case $\Omega_\Lambda = 0$ as
\begin{eqnarray}
\ell_1 \simeq 200 {\Omega_0}^{-0.5}
\label{f4.3}\end{eqnarray}
For the case $\Omega_m = 0.3$ and $\Omega_0$ near 1 Weinberg (2000) has 
derived the relation 
\begin{eqnarray}
\ell_1 \simeq 200 {\Omega_0}^{-1.58}.
\label{f4.4}\end{eqnarray}
Near $\Omega_0 = 1$ the relation (\ref{f4.3}) and  (\ref{f4.4}) are of course
very similar. Let us now turn to their results. BOOMERANG observes 
(de Bernardis et al. 2000) that the position of the first multipole 
peak occurs at $\ell_1 = 197\pm 6$ which corresponds to 
\begin{eqnarray}
\Omega_0 = (200/\ell_1 )^2 = 1.01\pm 0.02\ .
\label{f4.2}\end{eqnarray}
This value is very weakly dependent on a large number 
of parameters which mostly get determined by the shape of the multipole 
spectrum above the region of the first peak (Lange et al. 2000). Actually 
BOOMERANG and MAXIMA-1 fit their multiparameter data by the program 
CMBfast, so they do not explicitly use either (\ref{f4.3}) or (\ref{f4.4}).  
For the BOOMERANG results, there are no precise confidence levels published, 
only a coarsely pixelized likelihood surface of 95\% 
confidence. Other confidence contours such as 68\% CL, 90\% CL are not 
reported. In consequence we do not have sufficient information to use our 
polynomial fit to the data. Of course the maximum likelihood method can   
compute these confidence contours by using the 95\% CL information above, 
but this is not reliable at all. Therefore, we did not include the 
BOOMERANG data as an independent constraint, but it has been included 
together with the Large Scale Structure constraint (Novosyadlyj et al. 2000b). 

Now we turn to MAXIMA-1. From the statistical point of view this is less 
precise than that of BOOMERANG. They find the first acoustic peak  
at $\ell_1 \simeq 210$.
Therefore, we are not surprised about the report of Balbi et al. (2000) 
$\Omega_0 = 0.90\pm 0.15$, where the error corresponds to a 95\% 
confidence level. If we convert this to a 68\% error, then
$\Omega_0 = 0.90\pm 0.08$. The likelihood contours in the 
$\Omega_m, \Omega_\Lambda$ plane from MAXIMA-1 (Balbi et al. 2000) 
are available to us 68\% $(1\sigma)$, 95\% $(1.96\sigma)$ and 99\% 
$(2.58\sigma)$ confidence levels respectively.

There is a contradiction in the acoustic peaks 
between the two observations. There is no clear information 
from where the difference comes. Nevertheless, both teams are analysing   
additional data which may reduce the conflict of the region of power spectrum 
where further peaks are expected.
In the meantime, to adjust the peaks between these two experiments, a 
clever technique has been applied by Jaffe et al. (2000). Both teams  
BOOMERANG and MAXIMA-1 have a remarkable calibration uncertainty in their 
method. Jaffe et al. (2000) have used this uncertainty for the adjustment of
the peaks. The new acoustic peak is then between BOOMERANG and MAXIMA-1. 
MAXIMA-1 $\downarrow$ and BOOMERANG $\uparrow$ Fig. \ref{B+Mfig}. 

\begin{figure}[h]
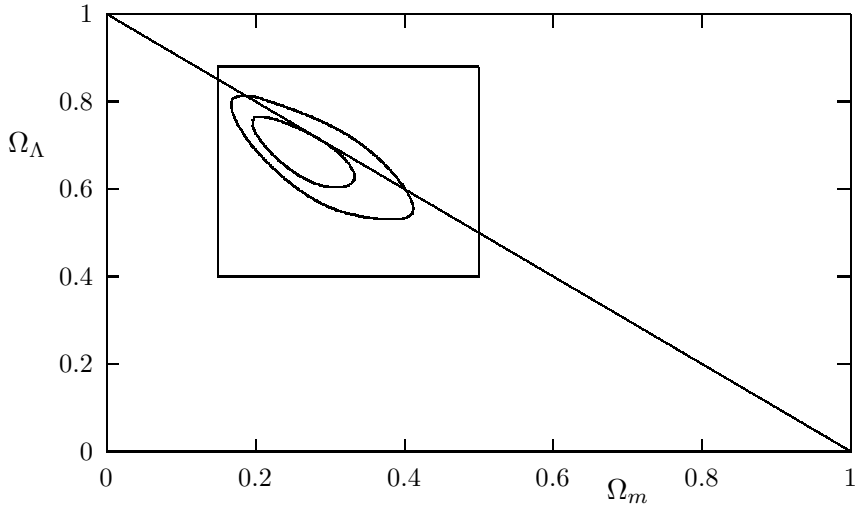

\centering
\input 1to5.tex
\caption{Confidence contours of the log-likelihood sum of two SN Ia 
(HSST and SCP), CMBR, Double Radio Galaxies and MAXIMA. The inner and outer 
curves correspond to 1$\sigma$ and 2$\sigma$  in the 
($\Omega_m, \ \Omega_{\Lambda}$)-plane, respectively. The significance of 
the square is described in the text.
The diagonal line corresponds to a flat cosmology.}
\label{fig4.5}
\end{figure}

Let us see our confidence region including balloon data. In Fig.
\ref{fig4.5} we show the confidence contours of the log-likelihood
sum of the SNe Ia, CMB, Double Radio Galaxies and MAXIMA-1.
We have not yet included BOOMERANG data in this fit. It will be included 
with LSS in the total fit. We have plotted our polynomial approximation only 
in the chosen window of $\Omega_m$ and $\Omega_{\Lambda}$.  

\section{Large Scale Structure}

Today's great challenge of science is to understand the structure 
of our Universe. One widely used theory on observational cosmology concerns  
structure formation yielding Large Scale Structures  
(LSS), and it was originated from the theory of Big Bang. The form of 
the power spectrum strongly depends on the cosmological parameters. 
From the observation of the power spectrum of density fluctuations one can 
apply a constraint on $\Omega_m$, $\Omega_\Lambda$. In a series of Papers  
Novosyadlyj et al. (1999, 2000a, 2000b, 2000c), Durrer and Novosyadlyj (2000) 
have analysed up to eight parameters in the $\chi^2$ fit to obtain 
the cosmological parameters. They have used the constraint on the amplitudes 
of power spectrum fluctuations from different sources on a wide scale. They 
have combined the power spectrum of density 
fluctuations of Abell-ACO clusters of Retzlaff et al. (1998) with optical 
determinations of the mass function of nearby galaxy clusters by Girardi et al. 
(1998), with evolution of the galaxy cluster X-ray temperature distribution 
function by Viana \& Liddle (1999a), Bahcall \& Fan (1998), with a study 
of bulk flows of galaxies by Kolatt \& Dekel (1997), and with Ly-$\alpha$ 
absorption lines 
in quasar by Gnedin (1998), Croft et al (1998). They apply a constraint 
on baryon density from big bang nucleosynthesis (BBN) by Burles et al. 
(1999), and the value of the Hubble constant 
$h = 0.65 \pm 0.10$. The Hubble value is low but reasonable within errors. 
The position of the first acoustic peak from the BOOMERANG observations,  
$\ell_1 = 197 \pm 6$, has also been included in their analysis.

There are seven continuous free parameters in this fit. They have fitted 
the parameters for a fixed neutrino number density ($N_\nu$ = 1, 2, and 3) 
in a mixed dark matter model with a cosmological constant ($\Lambda$MDM).
The $\chi^2$ is minimal for one species of massive neutrinos, and the
values of mass density and vacuum energy density are then 
$\Omega_m = 0.37_{-0.15}^{+0.25}$, $\Omega_\Lambda = 0.69_{-0.20}^{+0.15}$.
The likelihood contours in the 
$\Omega_m, \Omega_\Lambda$ plane from LSS (Novosyadlyj et al. 2000c) 
are available to us 68.3\% $(1\sigma)$, 95.4\% $(2\sigma)$ and 99.73\% 
$(3\sigma)$ confidence level, respectively. 

\begin{figure}[h]
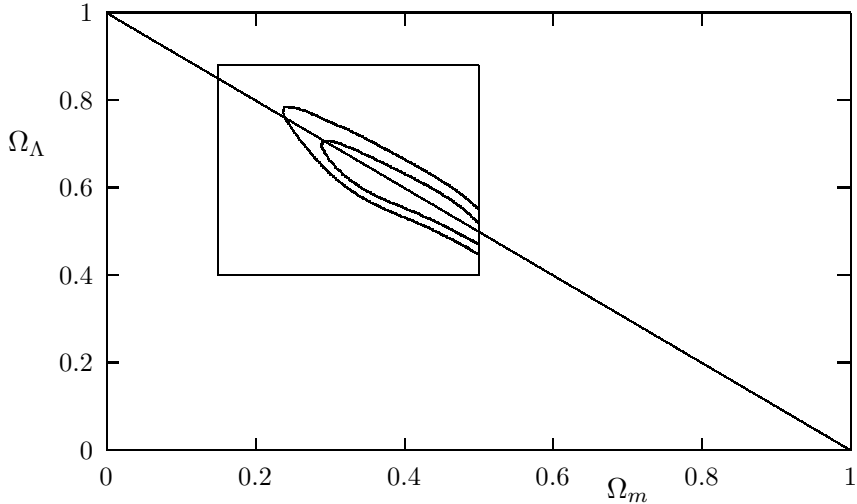

\centering
\input 4cons12.tex
\caption{The confidence contours of the log-likelihood sum of CMBR, 
MAXIMA-1, LSS and Double Radio Galaxies. The inner and outer curves 
correspond to 1$\sigma$ and 2$\sigma$ in the 
($\Omega_m, \ \Omega_{\Lambda}$)-plane, respectively. The significance 
of the square is described in the text. The diagonal line corresponds 
to a flat cosmology.}
\label{fig4.6}
\end{figure}

In Fig. \ref{fig4.6} we show the confidence contours of the log-likelihood
sum of CMB, MAXIMA-1, LSS and Double Radio Galaxies. 
We have plotted our polynomial approximation only in
the ranges of $\Omega_m$ and $\Omega_{\Lambda}$ that we sampled. 
As can be clearly
seen, the likelihood function contains information mainly on
$\Omega_0$, but it also gives a rather conspicuous upper limit on the
orthogonal combination $\Omega_{\Lambda} - \Omega_m$. The narrowness in this 
fit mainly comes from the CMB and MAXIMA-1. \\

\section{Fitting and Errors}

\begin{figure}[h]
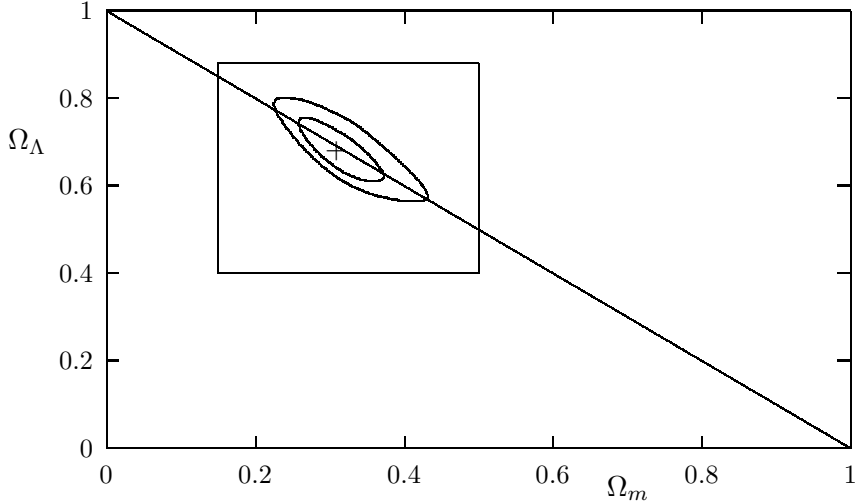

\centering
\input mlfit.tex
\caption{All the constraints described in this Chapter combined. 
The '+' marks the best fit: ($\Omega_m, \ \Omega_{\Lambda}$)
= (0.31,0.68). The inner and outer curves correspond to 
1$\sigma$ and 2$\sigma$ in the ($\Omega_m, \ \Omega_{\Lambda}$)-plane, 
respectively. The diagonal line corresponds to a flat cosmology.}
\label{fig7}
\end{figure}

Adding up our polynomial approximations to the confidence
contours of all the data fitted in this section, 
results in Fig. \ref{fig7}, where we show the location of the minimum, 
the $1\sigma$ and $2\sigma$ contours. From this Figure one can read
off the following results:
\begin{eqnarray}
\Omega_m = 0.31 ^{+0.12}_{-0.09}\\
\Omega_{\Lambda} = 0.68 \pm 0.12 ,
\label{f2}\end{eqnarray}
or alternatively
\begin{eqnarray}
\Omega_0 = 0.99 \pm 0.04 \\
\Omega_{\Lambda} - \Omega_m = 0.37 ^{+0.20}_{-0.23} .\label{f3}\end{eqnarray}

Of these results, only the determination of $\Omega_0$ is quite
precise and worth detailed attention. We can conclude from it
that a flat universe with $\Omega_0 = 1$ is very likely.

The error of $\pm 0.04$ is  mainly statistical, although it does
contain systematic errors due to mildly discordant experiments,
as discussed in the previous section. To this we have to add a
total systematic error which we evaluate as follows.

Before discussing the systematic errors
let us see the results taking into account the new result of lensing 
statistics Helbig (2000). As we know that there is a strong disagreement 
between lensing and supernovae data, it could be interesting just 
to see the combined results.  
We combine lensing data Helbig (2000) with fig. \ref{fig7}, the total 
results is then $\Omega_m = 0.34_{-0.03}^{+0.11}$, 
$\Omega_\Lambda = 0.63_{-0.10}^{+0.04}$ or alternatively 
$\Omega_0 = 0.97_{-0.04}^{+0.05}$
$\Omega_\Lambda - \Omega_m  = 0.29_{-0.18}^{+0.05}$. The best value is
excluded by the supernovae data at 1$\sigma \ {\rm CL}$, and it is excluded by 
lensing data at 97\% ${\rm CL}$. For this strong contradiction the nice 
shape of the contour plot is destroyed; that may be seen in the 
Paper IV fig. 4.

Perlmutter et al. (1998, 1999) have quoted a total systematic
error for $\Omega_m$ and $\Omega_{\Lambda}$ along the flat line of 
$\pm 0.05$. We
consider that the same error applies to the SN Ia data of Riess
et al. (1998), although they could not evaluate it from their
limited sample of SNe Ia. Displacing both the SN Ia contours by
$\pm 0.05$ along the flat line, we obtain a very small systematic
error in the $\Omega_0$ direction
\begin{eqnarray}
\Delta_1\Omega_0 = ^{+0.012}_{-0.006} . \label{f4}\end{eqnarray}

There are also two kinds of systematic errors inherent in our
method of analysis. Firstly, we are reading off the coordinates
of the confidence contours of the different experiments with some
finite precision. We estimate this to be
\begin{eqnarray}
\Delta_2\Omega_0 = 0.027 . \label{f5}\end{eqnarray}

Secondly, since we only use 20 points to fit the confidence
contours of each experiment, there is an arbitrariness in their
choice; all we require is that the confidence contours should be
well fitted by whichever polynomial. This polynomial
arbitrariness results in a systematic error estimated to be
\begin{eqnarray}
\Delta_3\Omega_0 = 0.01 . \label{f6}\end{eqnarray}

The quadratic sum of the errors in Eqs. (\ref{f4}), (\ref{f5}),
(\ref{f6}) is then
\begin{eqnarray}
\Delta_{tot}\Omega_0 = 0.03 . \label{f7}\end{eqnarray}

Thus our final result for $\Omega_0$ is
\begin{eqnarray}
\Omega_0 = 0.99 \pm 0.04 \pm 0.03 , \label{f8}\end{eqnarray} where
the first error is statistical and the second error is systematical.
Thus our total error is $\pm 0.05$.

Let us now turn to the case of exact flatness, 
$\Omega_m = 1 - \Omega_{\Lambda}$.
Along the flat line the SN Ia systematic error is somewhat larger
than in the case of Eq. (\ref{f4}),
\begin{eqnarray}
\Delta_1\Omega_0^{flat} = \pm 0.025 . \label{f9}\end{eqnarray}

Our result is then
\begin{eqnarray}
\Omega_m^{flat} = 0.31 \pm 0.04 \pm 0.04, \label{f10}\end{eqnarray}
where the first error is statistical and the second error is 
systematical. Thus our total error is $\pm 0.055$. 

The overall constraints on cosmological parameters in year 2001 
are summarized in table \ref{table4.3}

\begin{table}[h]
\begin{center} 
\begin{tabular}{clc}  
\hline
{\em Parameters}  & values & {\em References} \\ 
\hline
$\Omega_m$ & $0.31\pm{0.06}$  & Harun-or-Rashid \& Roos 2001 \\
$\Omega_\Lambda$ & $0.68\pm{0.07}$  & Harun-or-Rashid \& Roos 2001 \\
$\Omega_0$ & $0.99\pm{0.04}$ & Harun-or-Rashid \& Roos 2001 \\
$t_0$ & $13.5\pm{1.3}$ & Roos \& Harun-or-Rashid 2001 \\
$H_0$ & $73\pm{7}$ & Gibson \& Brook 2001  \\
$\Omega_{b}h^2$ & $0.020\pm 0.001$ & Burles et al. 2001 (BBN) \\
$\Omega_{b}h^2$ & $0.022_{-0.003}^{+0.004}$ & de Bernardis et al. 2001 (CMB) \\
$\alpha$ & -1 $<$ \ but $<$ - ${1\over 3}$ & Perlmutter et al. 1999 (SNe Ia) \\
$\alpha$ & -1 $\leq$ \ but $<$ - ${1\over 2}$ & Bludman \& Roos 2001 (Quintessence) \\

\hline  
\end{tabular}
\end{center}
\caption {Constraints on cosmological parameters. Errors are given at 
1$\sigma$ confidence level. The Hubble constant $H_0$ in units of 
$\rm{km \ s^{-1}Mpc^{-1}}$ and $t_0$ is in Gyr.}
 \label{table4.3}
\end{table}

\chapter{Search for Baryonic Dark Matter in Clusters}

\section{What is Dark Matter?} 

\noindent
Observations tell that more than 90\% matter in the 
Universe is dark, but it is mostly obscure to the physicist and astronomer.  
In general, any form of matter which exists in the Universe 
in a non-luminous form is called dark matter. Two types of dark matter exist,  
namely baryonic dark matter and non-baryonic 
dark matter. The main baryonic dark matter 
candidates are Massive Compact Halo Objects (MACHOs), white, red and brown 
dwarfs, planetary objects, neutron stars, black holes, dust clouds.  
The MACHO collaboration (Alcock et al. 1997, 2000) 
has surveyed the halos in the Large Magellanic Cloud (LMC) 
and the Small Magellanic Cloud (SMC) at different scales. 
Searching for microlensing events toward LMC  
Alcock et al. (2000) reported that the most likely MACHOs mass is between  
$0.15M_\odot$ and $0.9M_\odot$. Among known astrophysical objects white dwarfs 
is in this mass range, so they are good candidates for MACHOs.  
Most recently, a large number of dead stars have been detected at the edge 
of our galaxy, the Milky Way (Oppenheimer et al. 2001). 
Extensions of the Standard Model of Particle Physics provide 
non-baryonic candidates to dark matter. From the cosmological point of view 
two main categories of non-baryonic candidates have been proposed: Cold 
Dark Matter (CDM) and Hot Dark Matter (HDM) according to whether they were 
slow or fast moving at the time of the galaxy formation. The typical HDM 
candidates are neutrinos of a few eV, whereas in the CDM sector, typical  
candidates are heavy Dirac or Majorana neutrinos in the GeV-TeV mass range 
or Weakly Interacting Massive Particles (WIMPs) with unknown properties.   
Supersymmetric theories offer a number of candidates, such as  
neutralinos and axions. 

A good object for the study of 
baryonic dark matter in the Universe is a cluster of galaxies. 
The size of the clusters are big and the edges are not well defined. At 
large distance from the cluster center their properties approach average 
universal properties. For this analysis Advanced Satellite for Cosmology 
and Astrophysics (ASCA) data will be used from 
the six X-ray clusters A401, A496, A2029, A2199, A2256, A3571. A precise 
value of $\Omega_m$ can be used in different contexts. 
Our new result of $\Omega_m$ from the Chapters III and IV will be
used for the analysis of baryonic dark matter in these clusters. 
Note that this data set has also been analyzed by Nevalainen et al. 
(1999, 2000a, 2000b) for constraining cosmological parameters.
The measurement of gas fraction in these clusters resembles their analysis. 

\section{Clusters of Galaxies} 

\noindent
Clusters are composed of baryonic and non-baryonic matter. The baryonic 
matter takes the forms of hot gas emitting X-rays, stellar mass observed in 
visual light, and perhaps invisible baryonic dark matter of unknown 
composition. Let us denote the respective fractions as $f_{gas},\ f_{gal} $, 
$ f_{bdm} $. It has been observed in a large number of clusters that 
$f_{gas}$ 
is an increasing function of radius (e.g. White \& Fabian 1995; Markevitch 
et al. 1997, 1999; Ettori \& Fabian 1998; Nevalainen et al. 1999, 2000b), 
approaching the universal ratio at a large radius, as deduced from cluster 
formation simulations (e.g. White et al. 1993, Frenk et al. 1999, Tittley \& 
Couchman 2000). Denoting the universal baryonic density parameter as  
$\Omega_b$ and the universal total mass density parameter as $\Omega_m$, 
one can write
\begin{eqnarray}
f_{gas}(r) + f_{gal}(r) + f_{bdm}(r) = 
\Upsilon(r){\Omega_b\over\Omega_m}\ ,
\label{main}\end{eqnarray}
where $\Upsilon(r)$ describes the possible local enhancement or diminution of 
baryon matter density in a cluster, compared to the universal baryon density.
One can make different conclusions from the Eq. \ref{main}. Let us turn 
to known and unknown quantities in Eq. (\ref{main}). The value 
of $\Omega_m$ is precisely known (Paper IV) and  
$\Upsilon$ is known with error and there is a strong constraint on it 
from simulations by Frenk et al. (1999). If we trust observational data,   
$f_{gas}(r)$ is known from fig. \ref{fig8},  
the value of baryonic density $\Omega_b$ is also known but 
contradictory between BBN and CMB (as has been discussed in the previous 
Chapters). The unknown or poorly known quantity in the above expression 
is $f_{bdm}$. Until the contradiction is resolved, the baryonic density 
can be used from BBN and CMB independently.

It has been known before that the gas fractions in clusters of galaxies are 
too high for Eq. (\ref{main}) to be satisfied with $\Omega_{m} = 1$ 
(e.g. White et al. 1993; David et al. 1995; White \& Fabian 1995; Ettori \& 
Fabian 1998; Mohr et al. 1999). The qualitative conclusion given by the lower 
accepted limit for $\Omega_{m}$ is that a large fraction of baryonic 
matter other than gas is ruled out.
Using up-to-date data on $f_{gas}$, $\Omega_b$, $\Omega_m$, $\Upsilon$ and 
$H_{0}$ one can set a quantitative limit to the sum of 
$f_{gal} $ and $f_{bdm}$ 
using Eq. (\ref{main}). 

\section{Gas Fraction} 

\noindent
Usually the total masses and gas fractions have been determined using the 
hydrostatic equilibrium condition with an isothermal assumption for the 
cluster gas. Here we have abandoned the isothermality assumption and instead
derived the gas mass fraction $f_{gas}(r)$ from a sample of six clusters, 
whose total mass profiles have been determined 
using the gas temperature profiles 
observed with ASCA. For this derivation I am indebted to Jukka Nevalainen. 
This is an important improvement upon the earlier work, 
because the accuracy of the total mass within a given radius is approximately 
proportional to the accuracy of the gas temperature at that radius. The gas 
mass profiles have been obtained from the ROSAT imaging data (see Nevalainen 
et al. (2000a) and the references therein for the original hydrostatic mass 
analyses of the individual clusters in the sample).

\begin{figure*}[h]
\centering
\psfig{figure=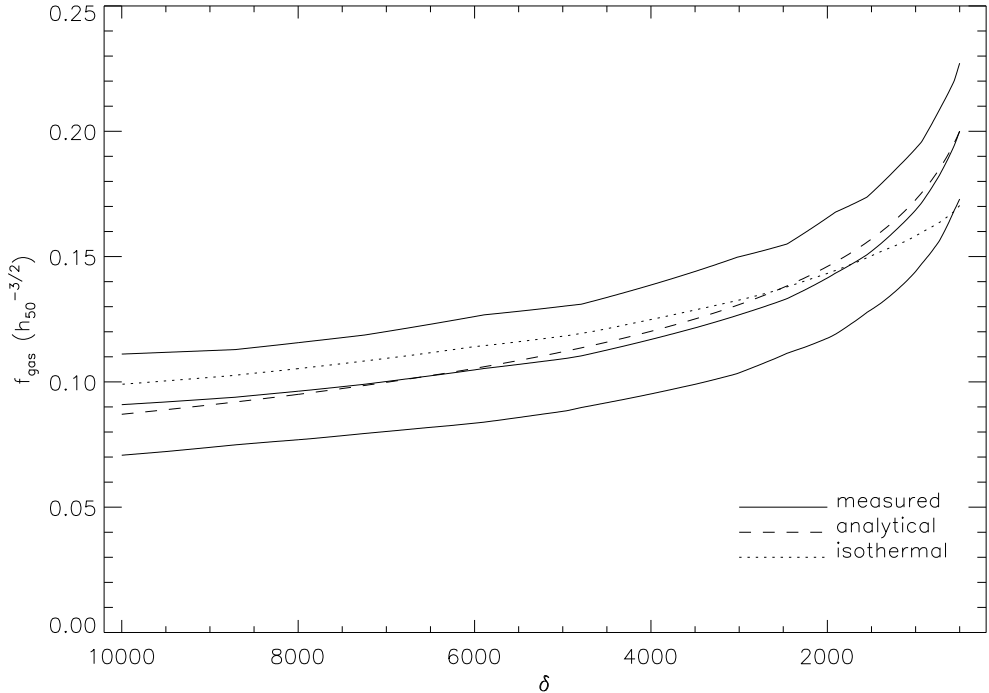,width=14cm,angle=0} 
\caption{The average $f_{gas}$ profile for the sample with 1 
$\sigma$ errors as a function of the overdensity $\delta$ (solid line). 
The analytical approximation is shown as a dashed line. 
The profile derived 
assuming gas isothermality is shown as a dotted line.}
\label{fig8}
\end{figure*}

We averaged the individual $f_{gas}(r)$ profiles without weighing them with 
their errors because the errors are not comparable due to different modeling 
of the temperature data in different cases. As the $1 \sigma$ error of the 
average we took the standard deviation of their distribution (the square root
of the unbiased variance of the data set). As a radial parameter we used the
overdensity, or the mean total mass density within a given radius in units 
of the critical density, $\delta(r) = \left<\rho(r)\right>/\rho_{c}$, where 
$\left<\rho(r)\right> = M_{tot}(<r)/(\frac{4}{3} \pi r^{3})$ and $\rho_{c} = 
3 H_{0}^{2}(1 + z)^{3}/8 \pi G$ ($z$ is the redshift of the cluster). 
Fig. \ref{fig8} shows that $f_{gas}$ increases with decreasing overdensity 
(or increasing radius) reaching a value
\begin{eqnarray}
f_{gas}(< r_{500}) = (0.200 \pm 0.027) h_{50}^{-3/2}
\label{gas}\end{eqnarray}
at a radius where the overdensity is 500. The radial increase is due to the 
steeper decrease of the dark matter density, compared to the gas density in 
these clusters. At the largest radii the $f_{gas}$ values increase rapidly, 
due to the decline of the gas temperature profiles. In the overdensity range 
[$10^{4}$ - 500] the measured $f_{gas}$ profile can be approximated by an 
analytical formula 
\begin{eqnarray}
f_{gas}(\delta) = [(\delta/500)^{-0.04} - 0.80] h_{50}^{-3/2}\ ,
\end{eqnarray}
where the relative error increases from 14\% at $r_{500}$ to 22\% at 
$r_{10000}$. The systematic uncertainties inherent in the hydrostatic mass 
determination method due to deviations from the hydrostatic equilibrium or 
spherical symmetry have been evaluated by simulations (e.g. Evrard et al. 
1996; Schindler 1996; Roettiger et al. 1996). The cluster sample used 
here has been selected for its lack of any signature for such deviations.  
These uncertainties are negligible compared to the above rms variation. 

Our result is consistent with isothermal analyses of cluster samples by  
Ettori \& Fabian (1998) and Mohr et al. (1999). For comparison, we also 
derived the average $f_{gas}$ profile for our sample deriving the total 
masses with hydrostatic equilibrium equation, assuming that the gas is 
isothermal. The resulting profile is shown in Fig. \ref{fig8}.
At small radii the isothermal $f_{gas}$ is higher than the measured value, 
and at large radii the opposite is the case. This is due to the fact that at 
small radii the measured temperatures and consequently 
the total masses are bigger 
than the isothermal values, but at the large radii the temperature profile 
drops below the average. Even though the isothermal profile is within 
$1 \sigma$ errors of the measured profile (the isothermal profile gives 
$f_{gas}(< r_{500}) = (0.170 \pm 0.025) h_{50}^{-3/2})$, the data indicates 
that at radii larger than $r_{500}$ the isothermal values would probably be 
significantly smaller than the measured values. 

Due to limitations of ASCA, the measured temperature profiles do not extend 
very far, and consequently the mass determinations are reliable only up to 
$\sim r_{500}$, at which radius we evaluate Eq. (\ref{main}). 


\section{Local vs. Universal Baryon Fraction} 

\noindent 
The cluster formation simulations give information on $\Upsilon(r)$, or how 
the  cluster baryon fractions relate to the universal baryon fraction. 
Frenk et al. (1999) have simulated the formation of an X-ray cluster in a 
cold dark matter universe (assuming $\Omega_m=1$, $\Omega_{\Lambda}=0)$ via 
hierarchical clustering using 12 different codes. Tittley \& Couchman (2000), 
tested the standard cold dark matter cluster formation via hierarchical and 
non-hierarchical clustering. Eke et al. (1998) simulated cluster formation 
in a flat, low density Universe ($\Omega_m=0.3$, $\Omega_{\Lambda}=0.7)$. 
All these simulations give very similar values for $\Upsilon(r)$ below 
unity, but approaching it at large radii. At the radii corresponding to 
$r_{500}$ in our cluster sample (1.3 - 1.7 $h_{50}^{-1}$ Mpc), the central 
68\% of the simulated values vary in the range
\begin{eqnarray}
\Upsilon = 0.9\pm 0.1\ .
\label{ups}\end{eqnarray}

\section{Baryonic Dark Matter} 

\noindent
From a sample of six clusters with measured temperature profiles we have  
determined an average $f_{gas}$ profile as a function of overdensity.
The measured $f_{gas}$ values are systematically, but not significantly, 
below the values determined with the usual assumption of gas isothermality 
at small radii, and above them at large radii.

We combine our result $f_{gas}(< r_{500}) = 0.200 \pm 0.027 h_{50}^{-3/2}$ 
with that of Burles et al. (2001) for the baryonic density parameter 
$\Omega_b$, with the mass density parameter $\Omega_m$ value from 
Paper IV, with the Gibson \& Brook (2001) value for 
the Hubble constant, and the Eke et al. (1998), Frenk et al. (1999) and 
Tittley \& Couchman (1999) value for $\Upsilon(r)$. We find the sum of 
$f_{gal}$ and $f_{bdm}$ is then 
\begin{eqnarray}
f_{gal}(<r_{500}) + f_{bdm}(<r_{500})=-0.010\pm 0.027\ . 
\label{central}\end{eqnarray}

Since the central value is in the unphysical region, it is more 
useful to express this result as an upper limit.  
In order to set a meaningful upper limit to the sum 
we must make use of the "unified" approach to classical statistics 
(Feldman \& Cousins 1998), which always produces confidence ranges 
entirely within the physical region. Then we obtain
\begin{eqnarray}
f_{gal}(<r_{500}) + f_{bdm}(<r_{500}) < 0.027\ \  (84\% {\rm CL}) \nonumber \\
f_{gal}(<r_{500}) + f_{bdm}(<r_{500}) < 0.045\ \  (95\% {\rm CL}).
\label{fin}\end{eqnarray}

Accepting all the input values, we conclude that the 
sum of the fractions of stellar matter $f_{gal}$ and baryonic dark matter 
$f_{bdm}$ within $r_{500}$ is very small, $< 0.027$, at 84\% {\rm CL}, 
$ < 0.045$,  at 95\% {\rm CL}. 
Our quantitative conclusion is in good agreement with qualitative results in 
earlier work. In particular, $f_{bdm}$ is of the order of a percent, 
unless the input parameters are wrongly estimated or conspire 
accidentally. But if $\Omega_b$ were taken to be as small as was 
estimated by Izotov et al. (1999) the situation would be 
further aggravated. Their estimation is 
$\Omega_bh_{100}^2 = 0.017 \pm 0.003$ 68\% CL), keeping all the values same,  
then the result is $f_{gal}(<r_{500}) + f_{bdm}(<r_{500}) = - 0.021 \pm 0.030$.
It is also reasonable to conclude an upper limit from this result, thus  
\begin{eqnarray}
f_{gal}(<r_{500}) + f_{bdm}(<r_{500}) < 0.029\ \  (84\% {\rm CL}) \nonumber \\
f_{gal}(<r_{500}) + f_{bdm}(<r_{500}) < 0.048\ \  (95\% {\rm CL}).
\label{f5.a}\end{eqnarray}

Let us use the baryonic density $\Omega_b$ from the recent balloon 
experiments, de Bernardis et al. (2000) and Balbi et al (2000),  
keeping all the other values same, then the result is $f_{gal}(<r_{500}) + 
f_{bdm}(<r_{500})= 0.066 \pm 0.046$. We can also conclude as a upper 
limit in the same way, that  
\begin{eqnarray}
f_{gal}(<r_{500}) + f_{bdm}(<r_{500}) < 0.049\ \  (84\% {\rm CL}) \nonumber \\
f_{gal}(<r_{500}) + f_{bdm}(<r_{500}) < 0.078\ \  (95\% {\rm CL}).
\label{f5.5}\end{eqnarray}

This value is rather high compared with the BBN case. Another discrepancy 
has been found Sadat \& Blanchard (2001) between observations and 
numerical simulations concerning $f_{gas}$. Simulation suggests that the 
baryonic fractions in X-ray clusters of galaxies are less than the   
observations found. 
In this situation our values in Eqs. \ref{fin} and \ref{f5.5}
will be increased. This discrepancy could be resolved from the precise  
data of X-ray Multi-Mirror Mission (XMM). However, it is clear that  
${\Omega_b}h^2$ needs to be revised. Most recently DASI (Pryke et al. 2001) 
and BOOMERANG (de Bernardis et al. 2001) have 
reported their ${\Omega_b}h^2$ values, which is a good agreement with BBN, 
their estimation is ${\Omega_b}h^2 = 0.022_{-0.003}^{+0.004}$ at 68\% CL.

\chapter{Black Hole Cosmology} 

\section{Spherical Black Holes}

\noindent
The theoretical idea of black holes is more than 200 years old, but due to 
the development of science and technology, black holes may become 
observable and interesting nowadays. 
New information about black holes is given by Chandra, who   
finds a more than 500 solar mass black hole in the M82 galaxy, which is 
located at a distance of 600 light years from the galactic center. 
This is the first confirmed case of such a large black hole outside the 
center of a galaxy (Chandra 2000, Kaaret et al. 2000, Matsushita et al. 2000). 

Recent evidence indicates that most of the galaxies have Super-massive 
Black Hole (SMBH) at their centers (Marconi et al. 2000, Peterson \& 
Wandel 2000). The formation of a black hole can follow different 
routes: i) A black hole is formed when a collapsed star has more 
mass than 3 solar masses. ii) A cluster of star-like black holes forms and 
eventually merges into a single black hole. iii) A single large gas 
cloud collapses to form a black hole. iv) A Red giant can explode into a 
supernova and become a black hole. In the case of SMBH, the second case is 
more likely. After the formation of a black hole, it starts to accrete 
matter around it Thorne (1994). It could be interesting to observe 
the accretion of dark matter (DM) in the SMBH. 

We consider that the velocity distribution of dark matter  $v_{DM}$ is 
isotropic. In the case of a non-rotating spherical black hole, $r_i=3r_c$, 
where $r_i$ is the gravitational radius and $r_c$ is the core radius. 
The angular momentum of the dark matter spiraling at $r_i$ is 
\begin{eqnarray}
\mathcal L_{DM} = m_{DM} \ v_{DM} \ r_i
\label{f61}\end{eqnarray}
where $m_{DM}$ is the mass of dark matter. 

\begin{figure}
\centering
\mbox{\epsfxsize=9.0truecm\epsfysize=5.0truecm\epsffile{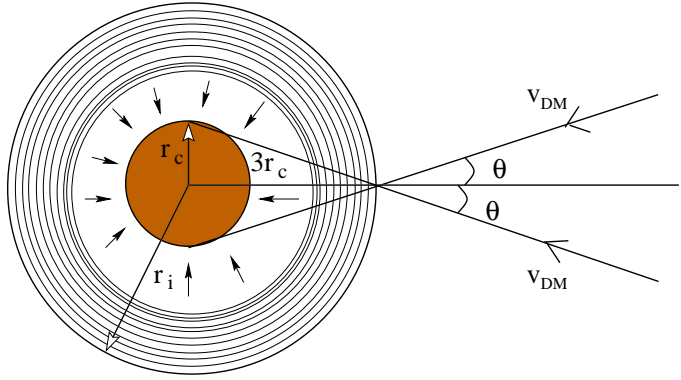}}
\caption{Geometry of a Super-massive Black Hole (SMBH). Particle spiraling 
and the condition that the particle falls into the hole are shown.}
\label{fig13}
\end{figure}

According to classical calculations, the condition that matter  
can spiral into the hole is that the rotational energy of matter should
be less than or equal to half of the gravitational potential energy 
(Longair 1994). This condition is fulfilled at radius $r_i$
\begin{eqnarray}
{\mathcal L_{DM}^2 \over 2I} \leq {GMm_{DM}\over 2r_i}
\label{f62}\end{eqnarray}
where $I$ is the moment of inertia $I=m_{DM}r_i^2$, and the condition that 
DM falls into the black hole is
\begin{eqnarray}
\mathcal L_{DM} \leq m_{DM}\sqrt{GMr_i}.
\label{f63}\end{eqnarray}
The rate of DM falling in the black hole is
\begin{eqnarray}
\dot{m}_{DM}= f \ \rho_{DM} \ v_{DM} \ 4\pi \ r_{i}^2
\label{f64}\end{eqnarray}
where $\it{f}$ is the fraction of particles falling into the hole. The 
above equation may be written in terms of the universal dark matter density 
$\Omega_{DM}$ and the Schwarzschild radius. After integrating we get
\begin{eqnarray}
M_0 - M_i = f[144 \ \pi \ \Omega_{DM} \ \rho_c \ v_{DM} \ G^2 \ t_o]{M_iM_0 \over c^4}
\label{f65}\end{eqnarray} 
where $M_i$ is the initial mass , $M_0$ is the present mass, and 
$t_0$ is the life time of the black hole. For $\it{f}$, a simple geometry 
may be considered, from Fig. \ref{fig13}  $v_\bot = v_{DM}sin\theta$. 
Using the angular momentum condition and replacing $c=v_{DM}$, one may 
apply a constraint on $\theta\leq 0.42$, leading to
\begin{eqnarray}
f\leq {2\theta\over \pi} \leq 0.27
\label{f66}\end{eqnarray}
Now all parameters are known and one can use Eq. (\ref{f65}) to determine 
the accretion of dark matter into a super-massive black hole in a fixed time. 
For instance if a SMBH of mass $10^6M_\odot$ accretes DM for 1 billion 
years then we get from Eq. (\ref {f65}) $M_0 - M_i \approx 10^{22}$ gm. 

According to theory, a black hole emits radiation like a black body. 
Let us see how fast it is evaporating. Collins et al. (1989) have shown that   
the time needed for a black hole to completely evaporate, i.e. to lose all 
its rest-mass energy $Mc^2$ through such radiation is  
\begin{eqnarray}
t \simeq {G^2 M^3 \over \hbar c^4},
\label{f6.a}\end{eqnarray}
where $\hbar$ is the Planck constant. The rate of evaporation is 
\begin{eqnarray}
{dM \over dt} = {\hbar c^4 \over 3 G^2 M^2}.
\label{f6.b}\end{eqnarray}
The evaporation depends on the mass of the black hole. So we can conclude 
that the evaporation is more effective in the case of a primordial black 
hole, and can be negligible for the SMBH case. 


\section{Time of Free Fall}
As we can see from the Fig. \ref{fig13} dark matter particles are rotating 
around a black hole, thus one can also calculate the rotational time of 
dark matter particles ($t_{DM}$) from the above geometry. Dark matter 
particles rotate around the black hole until they reach close to the core of 
the black hole. Particles gain kinetic energy as they approach toward 
the center, and are finally absorbed by the core. 
Particles can also directly fall at any time into the black 
hole if they lose the angular momentum on their way.  

The gravitational potential energy of a dark matter particle $m_{DM}$ at a 
radius $r_i$ is $U = -{GM m_{DM} \over r_i}$ and the corresponding force is 
$F = -{GM m_{DM} \over r_i^2}$, so ${dr \over dt} = -{GM t_{DM} \over r_i^2}$.
After integration and substituting the limits we get 
\begin{eqnarray}
{ 1 \over 3} (r_c^3 - r_i^3) = - {GM t_{DM} \over 2r_i}.
\label{f6.7}\end{eqnarray}
The gravitational radius $r_i$ is larger than the core radius $r_c$, 
i.e. $r_i^3$ $>>$ $r_c^3$. Neglecting the core radius $r_c$ we can write 
\begin{eqnarray}
t_{DM} = \sqrt{{2 \over 3} {r_i^3 \over GM}}.
\label{f6.8}\end{eqnarray}
Inserting the values of $G$, $M$ and $r_i$ we can obtain the rotational time 
of dark matter particles around a black hole. It is obvious from the Eq. 
(\ref{f6.8}) that the rotational time of dark matter particle depends on 
the gravitational radius of the black hole. 

\section{Toroidal Black Holes}

The mechanism of the high-energy jets from  
Active Galactic Nuclei (AGN) is not known. Here and in Paper V we study 
whether the geometry of a Toroidal 
Black Hole (TBH), in contrast to a Spherical Black Hole (SBH), 
could properly explain the particle dynamics of an AGN. 
From the mass accretion around a TBH, one 
can put a constraint on the life time of AGN activity (Paper V).  
A generalization of the black hole metrics can be given as follows
(Smith \& Mann 1997):
\[{ds}^{2}=-\left( V+b-\frac{2M}{R}\right) {dt}^{2}+\frac{{dr}^{2}}
{\left( V+b-\frac{2M}{R}\right)}\]
\begin{equation} 
+{R}^{2}[{d\theta }^{2}+c~{sinh}^{2}(\sqrt{a}\theta ){d\phi }^{2}],
\end{equation}
where $t$, $r$ are the time and radial coordinates, $\theta $ and $\phi $ are
coordinates on a two-surface of constant curvature and $V$ is a potential
term. An interpretation of the potential would be 
\begin{equation}
V=\frac{\Lambda}{3}{R}^{2},
\label{f6.2}\end{equation}
where $\Lambda $ is a negative constant in order to provide an anti 
de Sitter (AdS) spacetime.  
This looks like the cosmological constant, which however has 
a positive value (Perlmutter et al. 1997, 1999; Riess et al. 1998; 
Roos \& Harun-or-Rashid 2000), and which is a few orders of magnitude smaller
than needed to act as an effective AdS term. 
The parameters $b$, $c$, $a$
fix the topology of the structure: in particular, if $b=-a=0$ ($b=-a$ 
following the solution of Einstein field equations in empty space) and $a
\rightarrow 0$, $c=+\frac{1}{a}$, then the topology is that of a torus and
the space-time is asymptotically AdS. We recall that a
Schwarzschild metric for an asymptotically flat spacetime exhibits $b=0$ 
and +1 instead of the potential term $V$. Therefore, such a configuration 
is bounded to the presence of $V$. Some ideas about its origin will be 
discuss in the next section.  

\section{Lifetime of AGN Activity} 

\noindent
From the matter accretion 
around a TBH one can obtained the life time of AGN activity.  
The AGN activity is related to the toroidal shape of the black 
hole, so a transition to a quiescent state is expected as the black hole 
reaches SBH status. 
As matter accretes the hole, the event horizon increases,
the torus inflates and later it looses its starting configuration, turning
spherical. On the above line, we can infer a more quantitative estimate for
the lifetime of the activity phase by equating the metric tensor components
for a TBH and a SBH. If ${R}_{in}$ is an initial radial dimension 
for the torus, e.g. the middle value of the torus thickness with respect to 
the centre
of symmetry, and ${R}_{fin}$ is the final radius of the SBH, then the
transition happens if the following condition is matched:
\begin{equation}
V-\frac{2MG}{{R}_{in}{c}^{2}}=1-\frac{2MG}{{R}_{fin}{c}^{2}},
\end{equation}
or else:
\begin{equation}
{R}_{fin}=\frac{{r}_{g}}{1-V+\frac{{r}_{g}}{{R}_{in}}},
\end{equation}
where ${r}_{g}$ is the gravitational radius. Besides, the lifetime is related
to the accreted matter $\Delta M$ and to the accretion rate $\frac{dM}{dt}$
by:
\begin{equation}
dt(life)\sim \frac{\Delta M}{dM/dt},
\end{equation}
yielding:   
\begin{equation}
dt(life)\propto \frac{1}{dM/dt}{{R}_{fin}}^{3}.
\end{equation}
Since $\frac{dM}{dt}$ is observationally estimated,
we could enter the lifetime debate if only we knew the potential.
Lacking that, we can only study $dt(life)$ as a function of some
hypothetical functional form for it:
\begin{enumerate}
\item $V\simeq const$;
\item $V\propto 1/R$;
\item $V\propto log~R$;
\end{enumerate}
where the first case could be regarded as a cosmological vacuum energy
density, while the other expressions could mimic the background
potential of a surrounding axisymmetric galaxy. In particular, the second   
functional form refers to an embedding Newtonian gravitational field;
on the contrary, the logarithmic shape is motivated in order to
reproduce the flatness of the galactic rotation curve (Binney \&
Tremaine 1987). In this frame, the formation of a TBH can be deduced
if a protogalaxy develops a sufficiently extended background potential and the
collapse to a massive black hole is forced to lead to a toroidal
configuration.

\begin{figure}[h]
\centering
\epsfig{file=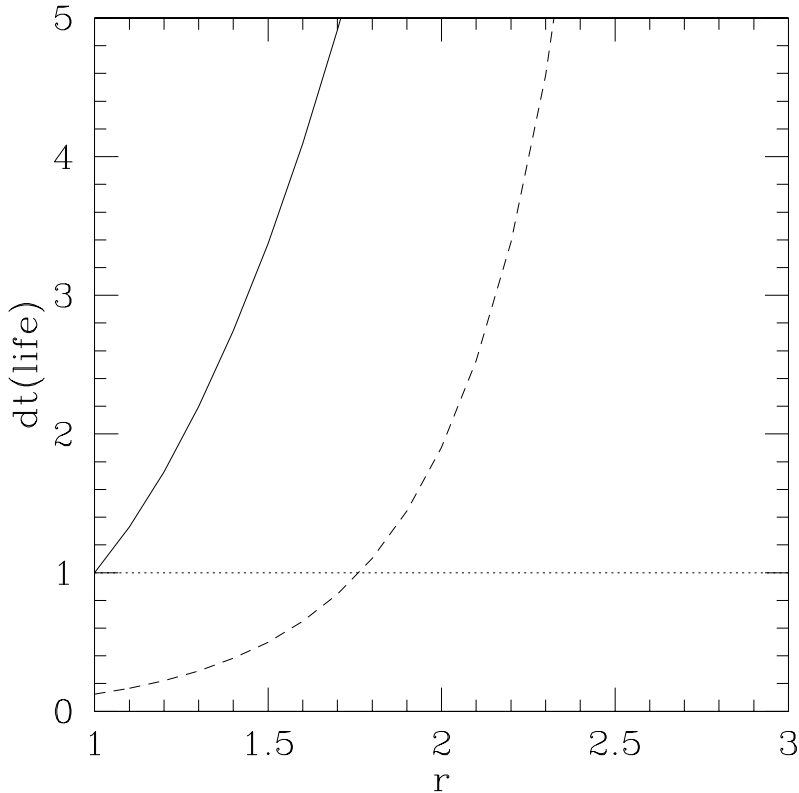,width=11cm}
\caption{AGN lifetime for a TBH model with constant accretion rate:
(\textit{solid line}) $V\simeq const$; (\textit{short-dashed line})
$V\propto 1/R$; (\textit{long-dashed line}) $V\propto log~R$ .
The normalization has been arbitrarily fixed and $r\equiv \frac{{R}_{in}}
{{r}_{g}}$.}
\label{fig14}
\end{figure}

In Fig. \ref{fig14}, the trend of $dt(life)$ is sketched for the above
mentioned different potentials, assuming a constant accretion rate $dM/dt$.
The $1/R$ potential gives a fixed value independent of ${R}_{in}$, while the
$V\simeq const$ situation gives an increasing function, as the   
$log~R$ potential does. If the AGN lifetimes
have small dispersion and are strictly focused on a single value, e.g.
${10}^{6}-{10}^{7}$ yr (e.g. Cavaliere \& Vittorini 2000), then the $1/R$
potential seems to give a likely interpretation to it.

Black hole mass is increasing rather than that of shrinking. In the case of a 
super-massive black hole Hawking radiation is negligible. Toroidal black 
hole is a nice tool for study AGN phenomena. From the mass accretion 
scenario in a Toroidal black hole, we can conclude that the lifetime of AGN 
activity is finite.

\chapter{Summary}

For the sake of minimum loss of experimental 
information and rigorous attention to statistics, both the least square 
method ($\chi^2$) and the maximum likelihood method have been applied to 
the analysis of a wide set of astrophysical data. 

The $\chi^2$ fit which 
uses 16 independent data constraints, all having published $1\sigma$ errors, 
gives the best value $\Omega_m = 0.33 \pm 0.07$ and  
$\Omega_\Lambda = 0.66 \pm 0.12$, or alternatively 
$\Omega_0 = \Omega_m + \Omega_\Lambda = 0.99 \pm 0.14$. In the exact flatness 
case $\Omega_m = 0.33 \pm 0.04$ and $\Omega_\Lambda = 0.67 \pm 0.04$ 
(Paper II, III). The goodness of fit of this 
analysis is extremely high, therefore we do not consider separately the 
systematic uncertainties in this fit.

The maximum likelihood method has been applied to 6 independent 
data, all having plotted contours of several different confidence levels, 
provides us the result  
$\Omega_m = 0.31_{-0.09}^{+0.12}$ and $\Omega_\Lambda = 0.68 \pm 0.12$,  
or alternatively $\Omega_0 = \Omega_m + \Omega_\Lambda = 0.99 \pm 0.04$. 
Including systematic uncertainties, the exact flatness case yields  
$\Omega_m = 0.31 \pm 0.04 (stat) \pm 0.04 (syst)$ and 
$\Omega_\Lambda = 0.69 \pm 0.04 (syst) \pm 0.04 (stat)$ (Paper IV). 
Since the results from both analyses are in good statistical agreement  
we have demonstrated that the least-squares method 
was correct and unbiased, and that the neglect of systematic  
uncertainties was of no importance.    

Thus both $\chi^2$ and 
maximum likelihood analyses confirm that the geometry of the Universe is 
flat, and the Einstein-de Sitter model with $\Omega_\Lambda = 0$, 
$\Omega_m = 1$ is strongly ruled out. Any low-density model with  
$\Omega_\Lambda = 0$ is ruled out. In addition, the age of the Universe is 
measured with a good precision to be $t_0 = 13.5 \pm 1.3 \ (0.68/h)$ Gyr,  
whereas $t_0 = 12.6 \pm 1.2 \ (0.73/h)$ Gyr.  
All errors are given at $1\sigma$ confidence level. 

From an analysis of six clusters of galaxies one concludes that  
clusters contain very little baryonic matter in excess of what is seen in 
X-rays. Making use of known 
values for the baryonic density parameter $\Omega_b$ and previously 
derived $\Omega_m$, one can establish a limit on the amount of dark 
baryonic matter in clusters. In order to get a meaningful result, the 
use of a special statistical technique is required.

Black hole masses increase due to the accretion of matter around 
it. In the case of super massive black holes, the evaporation rate is 
negligible compared to the accretion rate.  
A Toroidal Black Hole (TBH) study in contrast to Spherical Black
Holes (SBH) shows that the TBH can be used as an important 
tool in explaining AGN phenomena. The acceleration of particles,
production of jets, the shape of the magnetic field and the lifetime of
AGN activity are studied. 

Last few years physicists have been discussed the problem of cosmological  
constant with a great interest but the solution to this problem is still 
unresolved. A new model is needed to resolve this problem. Quintessence 
model has appeared in. However, we have to wait for the success of this model. 

The properties of dark matter in the Universe is mostly unknown, and 
the situation has remained unresolved since last twenty years. 
The dark matter problem can only be partly solved by the most recent 
discovery of dead stars in the Milky Way. A large portion is still
unresolved, solution of which would be a great success for astrophysics 
and cosmology.

\end{document}